\documentclass[prd,twocolumn,nopacs,floatfix,amsmath,nofootinbib,amssymb,floatfix]{revtex4}
\usepackage{graphicx,color,dcolumn,booktabs,bm}
\usepackage{longtable}
\usepackage{pdflscape}
\usepackage{pdfpages}
\usepackage{txfonts}
\usepackage{overpic}
\usepackage{amssymb}
\usepackage{makecell}
\usepackage{indentfirst}
\usepackage{feynmf}   
\usepackage{slashed}  
\usepackage{cases}
\usepackage{color}
\usepackage{multirow}
\usepackage{threeparttable}
\usepackage{epstopdf}
\usepackage{enumerate}
\usepackage{ulem}
\usepackage[figuresright]{rotating}
\usepackage[colorlinks,
            citecolor=blue,
            anchorcolor=red,
            menucolor=red,
            linkcolor=red,
            filecolor=red,
            runcolor=red,
            urlcolor=blue,
            frenchlinks=true]{hyperref}

\graphicspath{{Figures/}} %

\allowdisplaybreaks

\begin{document}
\title{Production of high-spin $\omega_J/\rho_J$ ($J=2,3,4,5$) mesons in $\pi^{-}p$ reactions}
\author{Ting-Yan Li$^{1,2,3,4}$}\email{lity2023@lzu.edu.cn}
\author{Zi-Yue Bai$^{1,2,3,4}$}\email{baiziyue@lzu.edu.cn}
\author{Xiang Liu$^{1,2,3,4}$}\email{xiangliu@lzu.edu.cn}
\affiliation{
$^1$School of Physical Science and Technology, Lanzhou University, Lanzhou 730000, China\\
$^2$Lanzhou Center for Theoretical Physics,
Key Laboratory of Theoretical Physics of Gansu Province,
Key Laboratory of Quantum Theory and Applications of MoE,
Gansu Provincial Research Center for Basic Disciplines of Quantum Physics, Lanzhou University, Lanzhou 730000, China\\
$^3$MoE Frontiers Science Center for Rare Isotopes, Lanzhou University, Lanzhou 730000, China\\
$^4$Research Center for Hadron and CSR Physics, Lanzhou University and Institute of Modern Physics of CAS, Lanzhou 730000, China}

\date{\today}
\begin{abstract}
In this work, we perform a comprehensive investigation of the production of high‑spin $\omega_J$ and $\rho_J$ mesons ($J=2,3,4,5$) in $\pi^- p$ reactions using an effective Lagrangian approach. By constructing the relevant $t$‑channel processes and calibrating the model with a single adjustable parameter fitted to existing data, we successfully reproduce the measured total and differential cross sections for the $J=3$ states $\omega_3(1670)$ and $\rho_3(1690)$.  Within the same framework, we predict the production cross sections for their lower- and higher-spin partners: $\omega_2(1975)$, $\rho_2(1940)$, $\omega_4(2250)$, $\rho_4(2230)$, $\omega_5(2350)$, and $\rho_5(2350)$. Our results show that these states exhibit measurable cross sections with characteristically forward‑peaked angular distributions, underscoring their strong potential for observation in future $\pi p$ meson‑beam experiments.
\end{abstract}

\maketitle

\section{INTRODUCTION} 

Over the past several decades, accelerator experiments have played an increasingly important role in exploring new phenomena in particle physics. However, when examining the status of light hadrons compiled by the Particle Data Group (PDG)~\cite{ParticleDataGroup:2024cfk}, it should be noted that the establishment of the light-hadron spectrum owes largely to meson-beam experiments (see review~\cite{WANG:2025fmh}). This historical fact underscores the continued relevance of $\pi p$ and $K p$ scattering studies today, even in the era of running experiments such as BESIII, Belle II, and LHCb. Building upon this perspective, numerous theoretical studies in recent years have focused on light hadrons produced in $\pi p$ and $Kp$ reactions \cite{Hou:2024ncf,Wang:2024qnk,Wang:2023lia,Wang:2022sib,Wang:2019uwk,Wang:2018mjz,Guo:2025ibo}. These investigations provide crucial information for elucidating the spectroscopy of such states, underscoring the importance of establishing and understanding these hadrons.

\begin{figure*}[htbp]
 \includegraphics[width=0.95
    \linewidth]{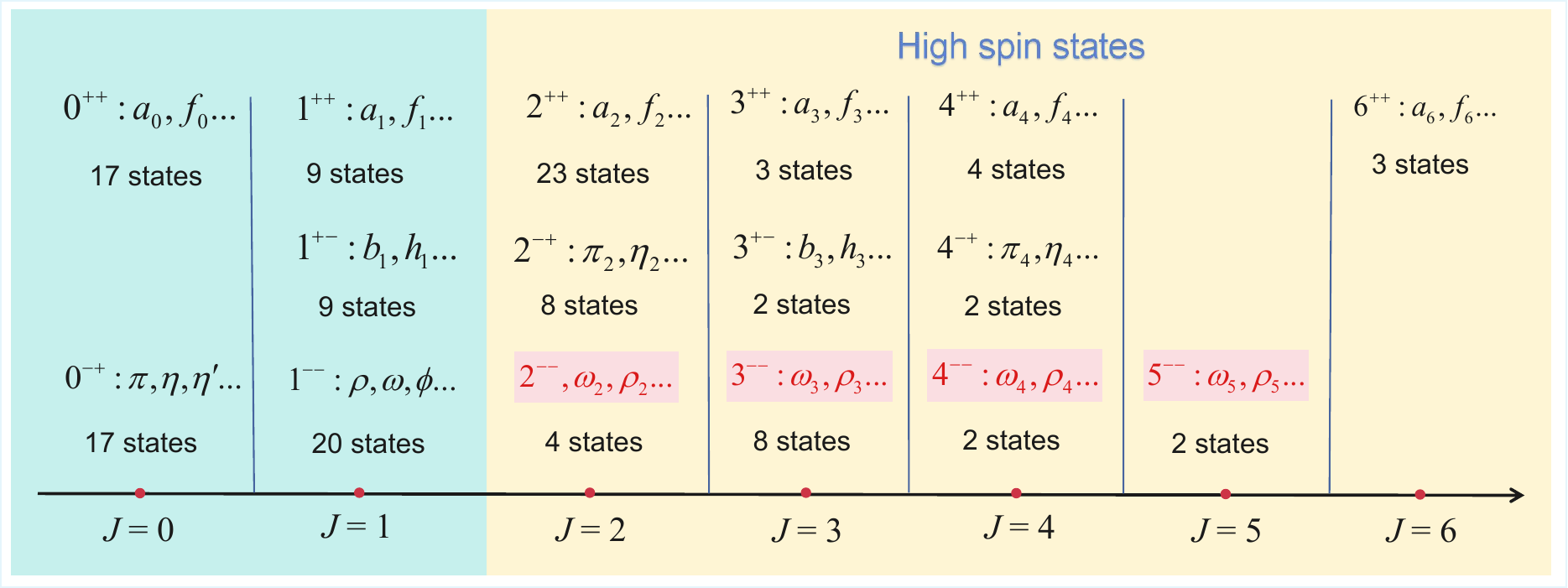}
\caption{A concise summary of light mesons \cite{ParticleDataGroup:2024cfk}.  }
\label{regge}
\end{figure*}

A particularly interesting subset of the light-hadron family is that of high-spin mesons, such as the $\rho_J$ and $\omega_J$ states with $J=2,3,4,5$. Their systematics can be summarized, as shown in Fig. \ref{regge}. However, precisely because of their high spin, detailed theoretical studies of these states remain relatively scarce \cite{Anisovich:2000kxa,He:2013ttg,Bai:2025knk,Li:2022khh,Wang:2022xxi,Pang:2015eha}. Their production dynamics are thus poorly understood, making them a pertinent target for investigation.

Meson-beam experiments are uniquely suited for this task. They offer a powerful tool for discovering and studying light hadrons, as evidenced by the numerous states observed in reactions like $\pi^- p$ scattering \cite{WANG:2025fmh}. Operating across a broad range of center-of-mass energies (from a few hundred MeV to several GeV), they probe hadrons with varying spins. The characteristically forward-peaked angular distributions in these processes further illuminate the underlying production mechanisms. These features make meson-beam facilities ideal for searching not only for the high-spin $\omega$ and $\rho$ states but also for other missing hadrons, keeping them indispensable in the ongoing effort to complete the hadron spectrum.

In this work, we investigate the production of high-spin $\omega_J$ and $\rho_J$ mesons using an effective Lagrangian approach combined with Reggeization. The study proceeds in two stages. First, we constrain our model parameters using available data for the production of the better-established $J=3$ states, $\omega_3(1670)$ and $\rho_3(1690)$, in $\pi^- p$ reactions. The sole free parameter is a universal cutoff, $\Lambda_t$, in the form factor.
Second, employing the same model and the extracted value of $\Lambda_t$, we present predictions for the total and differential cross sections of their lower- and higher-spin partners: $\omega_2(1975)$, $\rho_2(1940)$, $\omega_4(2250)$, $\rho_4(2230)$, $\omega_5(2350)$, and $\rho_5(2350)$, whose production has not yet been measured.
This approach provides a consistent, nearly parameter‑free link between the known $J=3$ sector and the unexplored $J=2,4,5$ sectors, offering concrete guidance for future experimental searches.

The paper is organized as follows. Following this Introduction, Section~\ref{section2} focuses on the production of $\omega_3(1670)$ and $\rho_3(1690)$ in $\pi p$ scattering, where the model is shown to describe the available experimental data well. Based on these results, Section~\ref{section3} presents predictions for the production characteristics of the $\omega$ and $\rho$ states with $J=2,4,5$. Finally, Section~\ref{section4} summarizes the work and its implications.

\section{Production of $\omega_3(1670)$ and $\rho_3(1690)$ in $\pi p$ reactions}
\label{section2}

In this section, we present a detailed study of the $\pi p$ scattering processes that produce the $J=3$ states $\omega_3(1670)$ and $\rho_3(1690)$. 

We begin by briefly summarizing the resonance parameters of the $\omega_3(1670)$ and $\rho_3(1690)$ in Table~\ref{mass}. According to the PDG~\cite{ParticleDataGroup:2024cfk}, the $\omega_3(1670)$ resonance was first observed in the $\pi^+ n \to p+3\pi$ process reported in Ref.~\cite{Armenise:1968hhd}. It is well established as a $1^{3}D_{3}$ state and its dominant decay channels include $\rho\pi$, $\rho\pi\pi$, and $b_1(1235)\pi$, as listed in the PDG.

The $\rho_3(1690)$ state, as the isovector partner of the $\omega_3(1670)$,  has been extensively studied in various production channels, such as $\pi^- p \to p+3\pi$, $\pi^- p \to K^+ K^- n$, $\pi^- p \to p+4\pi$, and $\pi^- p \to \eta\pi^-\pi^+ n$, among others~\cite{ParticleDataGroup:2024cfk}. Its decay modes are diverse, including $4\pi$, $\pi\pi$, $\omega\pi$, $a_2(1320)\pi$, $\rho\rho$, $\phi\pi$, $\eta\pi$, and several other final states.

\begin{table}[htbp]
\centering
\caption{The experimental information of the $\omega_3(1670)$ and $\rho_3(1690)$ states.
Here, the masses and widths are average values
taken from the PDG \cite{ParticleDataGroup:2024cfk}.  \label{mass}}
\renewcommand\arraystretch{1.5}
\begin{tabular*}{80mm}{@{\extracolsep{\fill}}lccc}
\toprule[1.pt]
\toprule[1.pt]
 $ I(J^{PC})$ & States & Mass (MeV) & Width  (MeV)\\
\hline
   $0~(3^{--})$      & $\omega_3(1670)$ &$1677\pm4$&$168\pm10$ \\
   $1~(3^{--})$      & $\rho_3(1690)$   &$1686\pm4$&$186\pm14$ \\
\bottomrule[1pt]
\bottomrule[1pt]
\end{tabular*}
\end{table}

Following the review of the experimental status of the $\omega_3(1670)$ and $\rho_3(1690)$ mesons, the quark pair creation (QPC) model is adopted to quantitatively calculate their dominant decay modes~\cite{Wang:2022xxi,Pang:2015eha}. The corresponding branching ratios obtained within the QPC model are listed in Table~\ref{decay}, which show good overall agreement with the available experimental data.

For the $\omega_3(1670)$, the dominant decay mode is $\pi\rho$ with a branching ratio of $0.66$, followed by significant contributions from $\pi b_1$ and $\eta\omega$. In contrast, the $K\bar{K}$ and $K\bar K^*$ channels are highly suppressed, each with a branching ratio of only $0.003$.

For the $\rho_3(1690)$, the largest decay channel is $\rho\rho$ with a branching ratio of $0.55$. Subleading contributions come from $\pi\pi$ and $\pi\omega$, while $\pi h_1$, $\pi a_2$, $\eta\rho$, $\pi a_1$, and $K\bar{K}$ exhibit smaller but non-negligible branching ratios.

\begin{figure}[tbp]
    \includegraphics[width=1\linewidth]{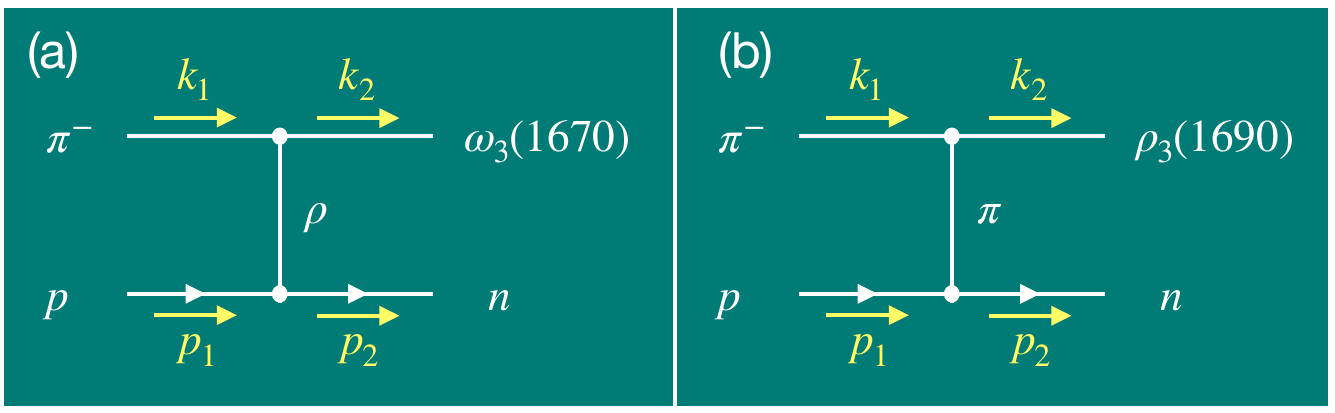}
\caption{ Feynman diagrams for the $\pi^{-} p\to  \omega_3(1670) n$ (a) and  $\pi^{-} p\to  \rho_3(1690) n$ (b) reactions.}
\label{Feynman diagram1}
\end{figure}

While the decay properties of these mesons have been studied theoretically, a systematic analysis of their production dynamics remains to be fully explored.  The $\pi^- p$ scattering process provides a suitable framework for this investigation, as both states have been historically observed in such reactions~\cite{Corden:1977xu,Engler:1974pd}. The feasibility of this approach is strongly supported by their decay patterns. Specifically, the $\omega_3(1670)$ decays predominantly via $\pi\rho$ (branching ratio $0.66$), while the $\rho_3(1690)$ has a significant $\pi\pi$ decay component. The prominence of these $\pi$-containing final states makes a description via $t$-channel meson exchange particularly natural. As shown in the tree-level Feynman diagrams in Fig.~\ref{Feynman diagram1}, the process $\pi^- p \to \omega_3(1670) n$ can proceed via $\rho$-meson exchange, and $\pi^- p \to \rho_3(1690) n$ via $\pi$-meson exchange.

\begin{table}[htbp]
\centering
\caption{The partial decay channels of the $\omega_3(1670)$ and $\rho_3(1690)$ states \cite{Wang:2022xxi,Pang:2015eha}. 
  \label{decay}}
\renewcommand\arraystretch{1.5}
\begin{tabular*}{80mm}{@{\extracolsep{\fill}}lccc}
\toprule[1.00pt]
\toprule[1.00pt]
 \multicolumn{2}{c}{$\omega_3(1670)$ } &   \multicolumn{2}{c}{$\rho_3(1690)$}   \\
 \hline
Channel  &{\it BR} &Channel  &{\it BR} \\
 \hline
$\pi \rho$  & 0.66 &$\rho \rho$  &0.55 \\
$\pi b_1$ &0.24  &$\pi \pi$  & 0.17\\
$\eta \omega$ &0.04  &$\pi \omega$ &0.12\\
$K\bar{K}$ & 0.003 &$\pi h_1$  & 0.06\\
$K\bar K^*$ & 0.003 &$\pi a_2$  & 0.04\\
   &  &$\eta \rho$  &0.02 \\
  &   &$\pi a_1$  &0.02 \\
  &   &$K\bar K$  &0.01\\
\bottomrule[1pt]
\bottomrule[1pt]
\end{tabular*}
\end{table}

In studying the reactions $\pi^{-} p \to \omega_3(1670)n$ and $\pi^{-} p \to \rho_3(1690)n$, we consider only the dominant $t$-channel contributions, as illustrated in Fig.~\ref{Feynman diagram1}. The $s$-channel contributions are highly suppressed because the intermediate nucleon pole mass lies far below the production threshold of the final-state $\omega_3/\rho_3$ and $n$. The $u$-channel contributions are also negligible, as the baryon–antibaryon decay modes of these states are kinematically disfavored or experimentally insignificant. We therefore employ an effective Lagrangian approach to calculate the corresponding $t$-channel amplitudes. 

\paragraph*{The reaction $\pi^{-} p \to \omega_3(1670)n$ }[Fig.~\ref{Feynman diagram1}(a)]. The reaction $\pi^{-} p \to \omega_3(1670)n$  proceeds dominantly via $t$-channel $\rho$-meson exchange. This mechanism is strongly supported by the decay properties of the $\omega_3(1670)$, in particular its dominant decay into $\pi\rho$ with a branching fraction of about $66\%$. Although the decay $\omega_3 \to \pi b_1$ is also allowed, the corresponding $t$-channel exchange of a $b_1$ meson gives a negligible contribution to the total cross section according our calculations. We consider the $W_3 V_1 P$ vertex  Lagrangian~\cite{Jafarzade:2021vhh},
\begin{equation}
\mathcal{L}_{W_3 V_1 P} = g_{W_3 V_1 P} \, \varepsilon^{\mu\nu\rho\sigma} \operatorname{tr}\!\left[W_{3,\mu\alpha\beta}\left\{\left(\partial_{\nu} V_{1,\rho}\right),\left(\partial^{\alpha}\partial^{\beta}\partial_{\sigma} P\right)\right\}_{+}\right],
\label{eq:omega3rhopi}
\end{equation}
where the fields $W_3$, $V_1$, and $P$ correspond to the spin-$3$ meson with $J^{PC}=3^{--}$, vector meson and pseudoscalar meson, respectively.
 The nonets of them read as
 $$ P=\frac{1}{\sqrt{2}}\left(\begin{array}{ccc}
\frac{\eta+\pi^{0}}{\sqrt{2}} &\pi^{+} &K^{+}\\
\pi^{-} &\frac{\eta-\pi^{0}}{\sqrt{2}} &K^{0}\\
K^{-} &\bar{K}^{0} &\eta
\end{array}\right), $$

$$ V_{1}^{\mu}=\frac{1}{\sqrt{2}}\left(\begin{array}{ccc}
\frac{\omega_{1}^{\mu}+\rho_{1}^{0\mu}}{\sqrt{2}} &\rho_{1}^{+\mu} &K_{1}^{*+\mu}\\[6pt]
\rho_{1}^{-\mu} &\frac{\omega_{1}^{\mu}-\rho_{1}^{0\mu}}{\sqrt{2}} &K_{1}^{*0\mu}\\[6pt]
K_{1}^{*-\mu} &\bar{K}_{1}^{*0\mu} &\omega_{1}^{\mu}
\end{array}\right), $$
$$
W_{3}^{\mu\nu\rho}=\frac{1}{\sqrt{2}}\left(\begin{array}{ccc}
\frac{\omega_{3}^{\mu\nu\rho}+\rho_{3}^{0\mu\nu\rho}}{\sqrt{2}} &\rho_{3}^{+\mu\nu\rho} &K_{3}^{+\mu\nu\rho}\\
\rho_{3}^{-\mu\nu\rho} &\frac{\omega_{3}^{\mu\nu\rho}-\rho_{3}^{0\mu\nu\rho}}{\sqrt{2}} &K_{3}^{0\mu\nu\rho}\\
K_{3}^{-\mu\nu\rho} &\bar{K}_{3}^{0\mu\nu\rho} &\omega_{3}^{\mu\nu\rho}
\end{array}\right).
$$
The specific Lagrangian we need is presented in detail as follows:
\begin{equation}
\mathcal{L}_{\omega_3 \rho \pi} = \frac{1}{2} g_{\omega_3 \rho \pi} \, \epsilon^{\mu \nu \rho \sigma} \, \omega_{3,\mu \alpha \beta} \, (\partial_\nu \rho_\rho^+) \, (\partial^\alpha \partial^\beta \partial_\sigma \pi^-),
\label{eq:omega3rhopi1}
\end{equation}
 fitting the partial width $\Gamma(\omega_3(1670) \to \rho\pi) = 168\ \mathrm{MeV}$~\cite{Pang:2015eha} yields the coupling constant $g_{\omega_3 \rho \pi} = 38.94\ \mathrm{GeV}^{-3}$. The $\rho NN$ vertex is given by~\cite{Nagels:2015lfa}
\begin{equation}
\mathcal{L}_{\rho N N}=  g_{\rho N N}\overline{N}\gamma_{\mu}N\rho^{\mu}+ \frac{f_{\rho N N}}{4m_p}\overline{N}\sigma_{\mu\nu}N(\partial^{\mu}\rho^{\nu}-\partial^{\nu}\rho^{\mu}),
\label{eq:rhoNN}
\end{equation}
where the values of $g_{\rho NN}$ and $f_{\rho NN}$ are listed in Table \ref{cp}.

\paragraph*{The reaction $\pi^{-} p \to \rho_3(1690) n$}  [Fig.~\ref{Feynman diagram1}(b)]. $\pi^{-} p \to \rho_3(1690) n$ is dominated by $t$-channel $\pi$-meson exchange. This choice is consistent with the decay pattern shown in Table~\ref{decay}, where the $\pi\pi$ mode constitutes the dominant $\pi$-containing decay channel. Contributions from exchanging isoscalar mesons (e.g., $\omega$ and $h_1$), which correspond to different isospin processes, are neglected in the present analysis based on the available data. We consider the $W_3 P P$ vertex Lagrangian~\cite{Jafarzade:2021vhh}: 
\begin{equation}
\mathcal{L}_{W_3 P P} = g_{W_3 P P} \operatorname{tr}\!\left[W_{3}^{\mu\nu\rho}\left[P,\left(\partial_{\mu}\partial_{\nu}\partial_{\rho}P\right)\right]_{-}\right],
\label{eq:rho3pipi}
\end{equation}
where the fields $W_3$ and $P$ correspond to the spin-$3$ meson with $J^{PC}=3^{--}$ and two pseudoscalar mesons. The detail $\rho_3 \pi \pi$ vertex is expanded as
\begin{equation}
\mathcal{L}_{\rho_3 \pi \pi} = \frac{1}{2} g_{\rho_3 \pi \pi} \, \rho_3^{0,\mu \nu \rho} \left[ \pi^+ \, \partial_\mu \partial_\nu \partial_\rho \pi^- - \pi^- \, (\partial_\mu \partial_\nu \partial_\rho \pi^+) \right].
\label{eq:rho3pipi1}
\end{equation}
Using the partial width $\Gamma(\rho_3(1690) \to \pi\pi) = 32\ \mathrm{MeV}$~\cite{Wang:2022xxi}, we obtain $g_{\rho_3 \pi \pi} = 12.02\ \mathrm{GeV}^{-2}$. The $\pi NN$ vertex is taken as~\cite{Nagels:2015lfa}
\begin{equation}
\mathcal{L}_{N N \pi} = \frac{f_{\pi NN}}{m_\pi} \, \bar{\psi}_N \gamma_\mu \gamma^5 \psi_N \, \partial^\mu \pi,
\end{equation}
with $f_{\pi NN}=0.95$.

\begin{figure}[htbp]
\includegraphics[width=7.8cm]{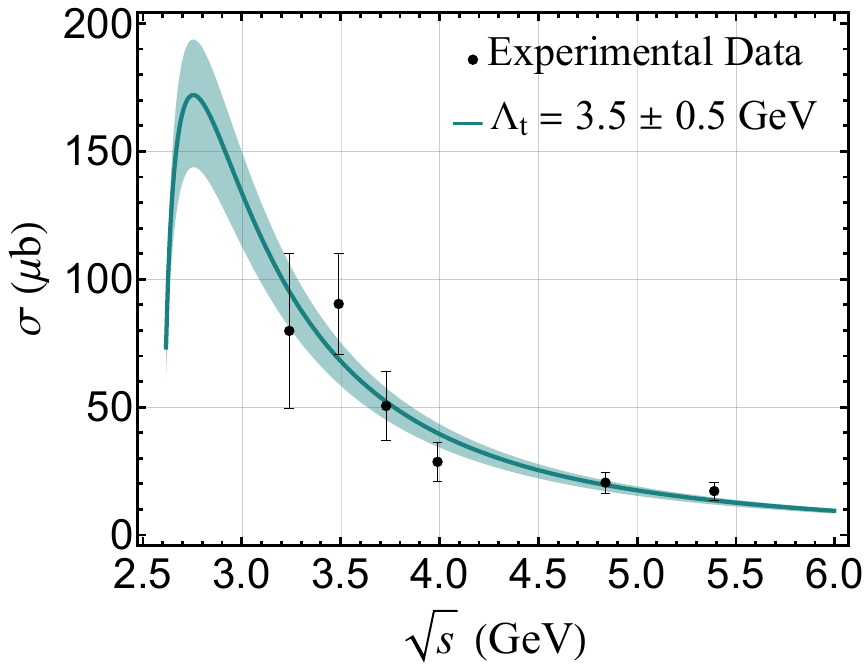}
\caption{The total cross section for the reaction $\pi^{-}p\to\omega_{3}(1670)n$. The black points with error bars correspond to the experimental data from Refs. \cite{Corden:1977xu,Diaz:1974bq,Matthews:1971br,Kenyon:1969ab,Armenise:1968hhd}.}
\label{omega3}
\end{figure}

\begin{figure}[htbp]
\includegraphics[width=8cm]{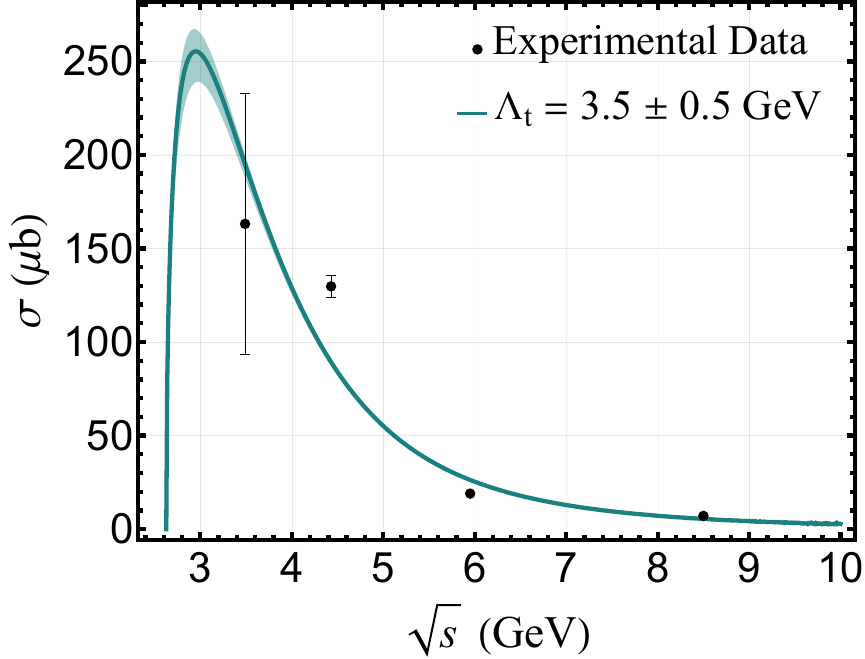}
\caption{The total cross section for the reaction  $\pi^- p \to \rho_3(1690)n$. The black points with error
bars correspond to the experimental data  from Refs. \cite{Engler:1974pd,BARI-BONN-CERN-GLASGOW-LIVERPOOL-MILAN-VIENNA:1980wqx,CERN-MUNICH:1975yng,GAMS:1994jop}.}
\label{rho3}
\end{figure}

With the above Lagrangians, the amplitudes for the production of $\omega_3(1670)$ and $\rho_3(1690)$ via \( t \)-channel \( \rho \) and $\pi$ exchanges in \( \pi^{-} p \) scattering can be written as:

\begin{equation}
\begin{split}
i\mathcal{M}_{\omega_3} = &\frac{1}{2} g_{ \omega_3 \rho \pi} \, \epsilon^{\mu\nu\rho\sigma} \,  q_{\nu} \, k^{\alpha}_1 \, k^{\beta}_1 \, k_{1 \sigma} \,\varepsilon_{\mu\alpha\beta}^*(k_2) \, \frac{-i \left( g^{\xi}_{\rho} + q^{\xi} q_{\rho} / m_{\rho}^{2}\right)}{t - m_\rho^2} \\
&\times \bar{u}(p_2) \left( g_{\rho N N}\gamma_{\xi} +\frac{f_{\rho N N}}{2m_p} \gamma_{\xi} \slashed{q}\right) u(p_1)F_t(q),
\end{split}
\label{f1}
\end{equation}

 \begin{equation}
\begin{split}
i\mathcal{M}_{\rho_3} = &-\frac{1}{2} g_{\rho_3\pi\pi} \frac{f_{\pi NN}}{m_\pi} \bar{u}(p_2) \gamma_\sigma \gamma_5 u(p_1) q^\sigma \varepsilon_{\mu\nu\rho}^*(k_2) \\
& \times (k_{1\mu} k_{1\nu} k_{1\rho} + q_{\mu} q_{\nu} q_{\rho}) \frac{i}{t - m_\pi^2} F_t(q),
\end{split}
\label{f4}
\end{equation}
where $\varepsilon(k_2)$ with different index denotes the polarization vector for the $\omega_3(1670)$ and $\rho_3(1690)$. $\bar{u}(p_2)$
 and $u(p_1)$ is the Dirac spinors of the nucleons. As detailed in Appendix \ref{APPENDIX B}, we present the complete summation formalism for the polarization tensors of the high-spin mesons with $J=3,4,5$. The $t$ is defined as the squared four-momentum transfer, $t = q^{2} = \left(k_{1} - k_{2}\right)^{2}$. For the $t-$channel meson exchange, the form factor  is adopted as
\begin{equation}
\begin{split}
F_{t}(q) = \left( \frac{\Lambda_{t}^{2} - m^{2}}{\Lambda_{t}^{2} - q^{2}} \right)^{2}, \end{split}
\label{Ft}
\end{equation}
where $m$ represent the mass of the exchanged particle. 
In this study, we take into account exclusively the $t$-channel contribution, so that the cutoff $\Lambda_t$ remains the sole adjustable parameter. 
By fitting the available experimental cross section data at different energies for the reaction $\pi^- p \to \omega_3(1670)n$~\cite{Corden:1977xu,Diaz:1974bq,Matthews:1971br,Kenyon:1969ab,Armenise:1968hhd}, we obtain $\Lambda_t = 3.5 \pm 0.5$ GeV, with $\chi^2/\mathrm{d.o.f.} \leq 1.3$.
Employing the same cutoff value $\Lambda_t = 3.5 \pm 0.5$ GeV for the $\pi^- p \to \rho_3(1690) n$ reaction, we find that the theoretical predictions are in good agreement with the measurements reported in Refs.~\cite{Engler:1974pd,BARI-BONN-CERN-GLASGOW-LIVERPOOL-MILAN-VIENNA:1980wqx,CERN-MUNICH:1975yng,GAMS:1994jop}. The results presented in Figs. \ref{omega3} and \ref{rho3}, show the total cross section for the $\pi^- p \to \omega_3(1670)n$ and $\pi^- p \to \rho_3(1690) n$ reactions calculated using $\Lambda_t = 3.5 \pm 0.5$ GeV.
For the remaining processes, namely $\pi^- p \to \omega_4(2250) n$, $\pi^- p \to \omega_5(2250) n$, 
$\pi^- p \to \rho_4(2230) n$, $\pi^- p \to \rho_5(2350) p$,
we adopt the identical cutoff $\Lambda_t = 3.5 \pm 0.5$ GeV throughout, to maintain consistency across the analyses.

The Regge trajectory model has been successfully applied to describe hadron production in the high-energy region~\cite{Ozaki:2009wp,Storrow:1983ct}. {The motivation for replacing the usual Feynman propagator with a Regge propagator is to effectively account for the exchange of high-spin and high-mass states in the $t$ channel, whose contributions cannot be neglected at higher energies.}
When employing conventional Feynman propagators, the resulting cross sections show a monotonically increasing dependence on the center-of-mass energy, which is inconsistent with the corresponding experimental measurements shown in Figs. \ref{omega3} and \ref{rho3}.
To address this problem, in the present framework, Reggeization is implemented by replacing the $t$-channel Feynman propagators in the Feynman amplitudes Eqs.~\eqref{f1} and \eqref{f4} with the corresponding Regge propagators  {\cite{Guidal:1997hy}}:
\begin{equation}
\begin{split}
\frac{1}{t-m_{\rho}^2} \rightarrow \left(\frac{s}{s_{\text{scale}}}\right)^{\alpha_{\rho}(t)-1} 
\frac{\pi\alpha_{\rho}^{\prime}}{\Gamma[\alpha_{\rho}(t)] \sin[\pi\alpha_{\rho}(t)]},
\end{split}
\label{rho_propagator}
\end{equation}
\begin{equation}
\begin{split}
\frac{1}{t-m_{\pi}^2} \rightarrow \left(\frac{s}{s_{\text{scale}}}\right)^{\alpha_{\pi}(t)}
\frac{\pi\alpha_{\pi}^{\prime}}{\Gamma[\alpha_{\pi}(t)+1] \sin[\pi\alpha_{\pi}(t)]},
\end{split}
\label{pi_propagator}
\end{equation}
where the scale parameter $s_{\text{scale}}$ is fixed at 1 GeV, {The functions $\alpha_\rho(t)$ and $\alpha_\pi(t)$ denote the $(J,M^2)$ Regge trajectories~\cite{Guidal:1997hy,Wang:2025rvr}:}
\begin{equation}
\begin{split}
\alpha_{\rho}(t) = 1 +\alpha_\rho^\prime(t-m_{\rho}^2),
\end{split}
\label{alpha_rho}
\end{equation}
\begin{equation}
\begin{split}
\alpha_{\pi}(t) = \alpha_\pi^\prime(t-m_{\pi}^2).
\end{split}
\label{alpha_pi}
\end{equation}
{where $\alpha_\rho^\prime = 0.8\,\text{GeV}^{-2}$ and $\alpha_\pi^\prime = 0.7\,\text{GeV}^{-2}$ are the slopes of the corresponding Regge trajectories, which can be generally constrained by the experimentally established states lying on the same trajectories \cite{Guidal:1997hy,Wang:2025rvr}.}

Based on the preceding theoretical framework, the cross section for the the reaction $\pi^{-} p\to \omega_3(1670)n$ and $\pi^{-} p\to \rho_3(1690)n$ can be calculated. The differential cross section in the center-of-mass (c.m.) frame is given by

\begin{equation}
\begin{split}
\frac{d\sigma}{d\cos\theta} = \frac{1}{32\pi s} \frac{\left|\vec{k}_{2}^{\,\mathrm{c.m.}}\right|}{\left|\vec{k}_{1}^{\,\mathrm{c.m.}}\right|} \left( \frac{1}{2} \sum_{\lambda} \left|\mathcal{M}\right|^{2} \right),
\end{split}
\label{diff}
\end{equation}
where $s = (k_{1} + p_{1})^{2}$ is defined, and $\theta$ is the angle between the outgoing meson and the incident $\pi^{-}$ beam direction in the c.m. frame. The three-momenta $\vec{k}_{1}^{\,\mathrm{c.m.}}$ and $\vec{k}_{2}^{\,\mathrm{c.m.}}$ correspond to the initial $\pi^{-}$ beam and the final meson, respectively. The factor of $1/2$ accounts for the average over the initial nucleon spin states.

 \begin{figure}[htbp]
\includegraphics[width=8.1cm]{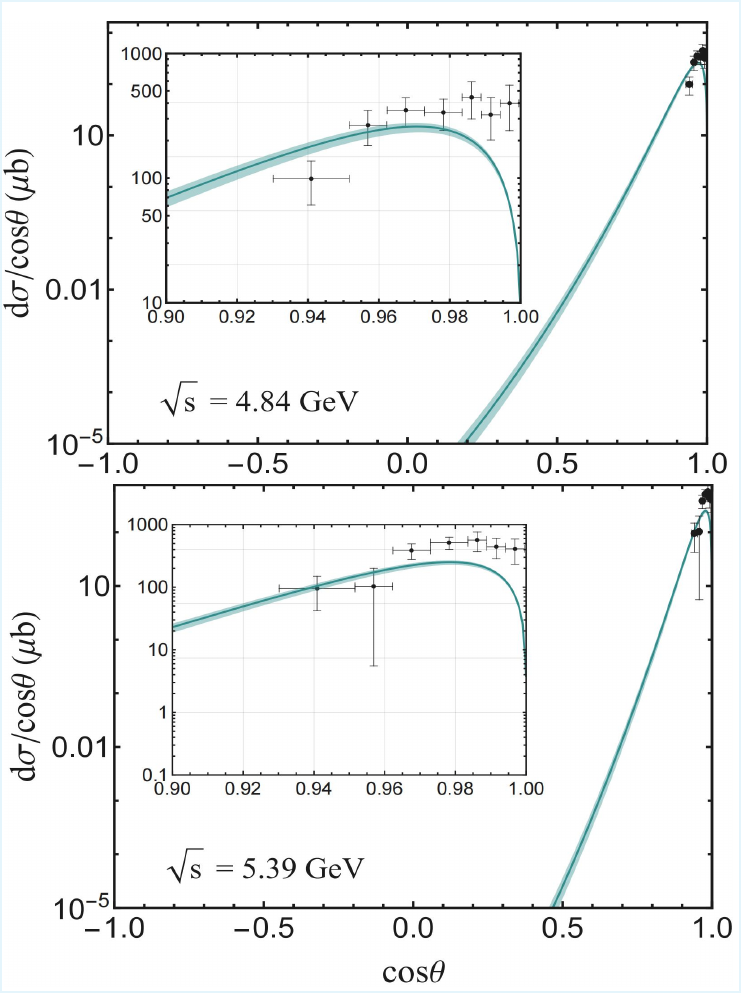}
\caption{  The differential cross section $d\sigma/d\cos\theta$
 of the $\omega_3(1670)$ production at different c.m. energy $E_{cm}$. The black points with error
bars correspond to the experimental data  from Ref. \cite{Corden:1977xu}.}
\label{omega3cs}
\end{figure}

\textbf{Numerical results.} The total cross section for the reaction $\pi^- p \to \omega_3(1670) n$ is presented in Fig.~\ref{omega3}. The experimental data (black points with error bars) are compiled from Refs.~\cite{Corden:1977xu, Diaz:1974bq, Matthews:1971br, Kenyon:1969ab, Armenise:1968hhd}. The theoretical calculation, with a cutoff parameter $\Lambda_t = 3.5\ \mathrm{GeV}$, is shown as the green solid curve. The associated uncertainty, estimated by varying $\Lambda_t$ by $\pm 0.5\ \mathrm{GeV}$, is represented by the green band. The theoretical curve closely follows the overall trend of the data across the measured energy range, demonstrating good agreement.

Among the data points in Fig.~\ref{omega3}, differential cross-section measurements are available only for the last two (highest) center-of-mass energies. To examine the angular distributions, the differential cross sections $d\sigma/d\cos\theta$ at $E_{\mathrm{cm}} = 4.84$ GeV and $5.39$ GeV are presented in Fig.~\ref{omega3cs}, together with the corresponding experimental data from Ref.~\cite{Corden:1977xu}. The data are concentrated at forward angles ($\cos\theta$ close to 1). For clearer comparison, subplots focusing on the region $0.90 < \cos\theta < 1.00$ are included in Fig.~\ref{omega3cs}.

At $E_{\mathrm{cm}} = 4.84\ \mathrm{GeV}$, the theoretical prediction (green solid curve) agrees excellently with the data over the full angular range, accurately reproducing a characteristic dip near $\cos\theta \approx 0.98$. At $E_{\mathrm{cm}} = 5.39\ \mathrm{GeV}$, slight deviations appear at very forward angles (around $\cos\theta \approx 0.97$). Nevertheless, the overall shape of the angular distribution, particularly its pronounced forward peaking, is well reproduced by the calculation. Although not all data points are encompassed at this higher energy, the theoretical curve captures the essential features of the measured distribution.

In summary, using the cutoff $\Lambda_t = 3.5\pm0.5\ \mathrm{GeV}$, our calculations provide a satisfactory description of both the total and differential cross sections for $\pi^-p\to \omega_3(1670)n$. This agreement supports the adequacy of the chosen parameter for modeling $\omega_3(1670)$ production in this reaction.

\begin{figure}[tbp]
\includegraphics[width=7.9cm]{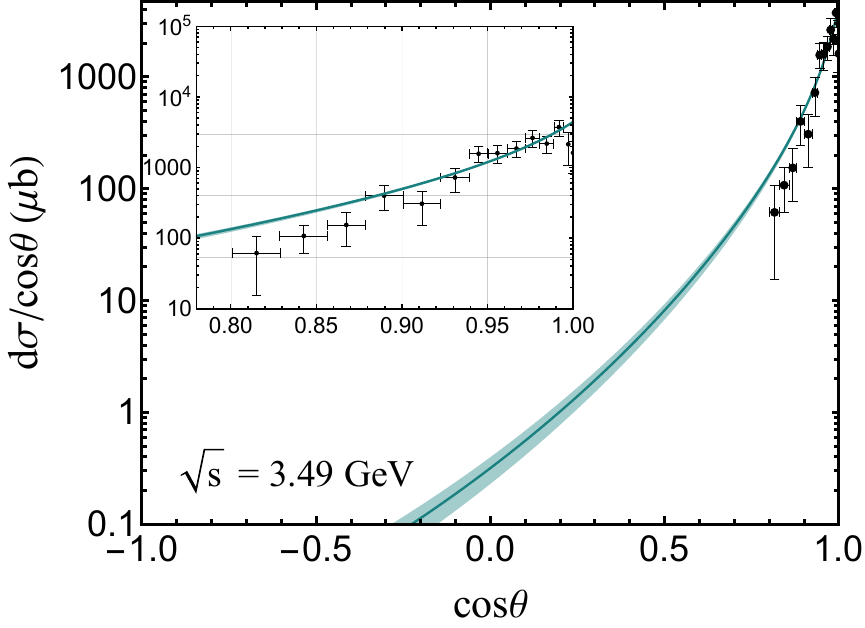}
\caption{  The differential cross section $d\sigma/d\cos\theta$
 of the $\rho_3(1690)$ production at c.m. energy $E_{cm}=3.49$ GeV. The black points with error
bars correspond to the experimental data  from Ref. \cite{Engler:1974pd}.}
\label{rho3cs}
\end{figure}

The production of the $\rho_3(1690)$ state in $\pi^- p$ scattering is analyzed within the same theoretical model used for the $\omega_3(1670)$. For consistency, we adopt the same cutoff parameter $\Lambda_t = 3.5\ \mathrm{GeV}$ and its associated uncertainty $\pm 0.5\ \mathrm{GeV}$ to estimate the theoretical band, as shown in Figs.~\ref{rho3} and \ref{rho3cs}.

Figure~\ref{rho3} presents the total cross section for $\pi^- p \to \rho_3(1690) n$. The experimental data from Refs.~\cite{Engler:1974pd, BARI-BONN-CERN-GLASGOW-LIVERPOOL-MILAN-VIENNA:1980wqx, CERN-MUNICH:1975yng, GAMS:1994jop} (black points with error bars) are well described within the theoretical uncertainty band across the displayed energy range. The theoretical curve reproduces the trend of the data, the cross section rises to a maximum around $\sqrt{s} \approx 2.9\ \mathrm{GeV}$ and then gradually decreases as the energy increases.

A more detailed angular analysis is provided in Fig.~\ref{rho3cs}, which shows the differential cross section $d\sigma/d\cos\theta$ at $\sqrt{s} = 3.49\ \mathrm{GeV}$. The theoretical prediction (solid green curve) and its associated uncertainty band ( corresponding to $\Lambda_t = 3.5 \pm 0.5\ \mathrm{GeV}$) are compared with the experimental data from Ref.~\cite{Engler:1974pd}. The predicted angular distribution exhibits a pronounced forward peak ($\cos\theta \to 1$). Excellent agreement with the data is achieved in this forward region, as clearly shown in the magnified inset.

With the parameter choice of $\Lambda_t = 3.5 \pm 0.5\ \mathrm{GeV}$, the model yields a consistent and satisfactory description of both the total and differential cross sections for $\omega_3(1670)$ and $\rho_3(1690)$ production. This agreement supports the validity of the adopted theoretical framework and indicates its reliability in capturing the production dynamics of these two $J=3$ mesons, which share the same $J^{PC}$ quantum numbers but differ in isospin.

{We note that the structures in the effective Lagrangians for high-spin fields are not uniquely determined. In this work, we adopt the minimal derivative rule: for each vertex, we retain the lowest number of derivatives consistent with Lorentz invariance and parity conservation. The coupling constants are fixed by the corresponding partial decay widths. Hence, any alternative Lagrangian form that reproduces the same width would yield a proportionally adjusted coupling, leaving the overall scale of the cross section largely unchanged. 
The validity of our minimal choice at low energies is explicitly verified by the successful fit to the $J=3$ total and differential cross section data.}

{At high energies, the Reggeization of $t$-channel propagators further reduces the sensitivity to vertex details. The Regge amplitude is dominated by the universal Regge trajectory $\alpha(t)$, which governs the energy dependence, while the vertex structure contributes only to the residue. 
Consequently, the predicted production features for the $J=4,5$ states are robust against moderate variations in the Lagrangian construction, as shown in the following section.}

\section{Production of higher-spin $\omega_J/\rho_J$ ($J=4,5$) mesons in $\pi^{-}p$ reactions}
\label{section3}

{We now extend the same theoretical framework to predict the production cross sections of the lower-spin and higher-spin $\omega_J$ and $\rho_J$ mesons ($J=2,4,5$) in $\pi^- p$ reactions.} This section is organized as follows. After a brief review of the experimental status of these states, we outline the theoretical formalism, including the effective Lagrangians and production amplitudes. Finally, we present the predicted total and differential cross sections.

{The resonance parameters for the $\omega_J$ and $\rho_J$ mesons with $J=2,3,4$ and $5$ are summarized in Table~\ref{mass2}. These states are the natural lower-spin and higher-spin partners of the $J=3$ $\omega_3(1670)$ and $\rho_3(1690)$. According to the PDG~\cite{ParticleDataGroup:2024cfk}, the $\omega_2(1975)$, $\rho_2(1940)$, $\omega_4(2250)$, $\rho_4(2230)$, and $\omega_5(2250)$ are categorized as ``further states,'' requiring more confirmation, whereas the $\rho_5(2350)$ is more firmly established.}

{Experimentally, these states have been observed in specific production channels.  The $\omega_2(1975)$ and $\omega_4(2250)$ are seen in $p\bar{p}$ annihilation, e.g., $p \bar{p} \to \omega\eta$ and $p \bar{p} \to \omega \pi^0 \pi^0$~\cite{ Anisovich:2002xoo}. Similarly, the $\rho_2(1940)$ and $\rho_4(2230)$ are found in $p\bar{p} \to \omega \pi^0$, $p\bar{p} \to \omega \eta \pi^0$, and $p\bar{p} \to \pi^+ \pi^-$~\cite{Anisovich:2002su}.} The $\omega_5(2250)$ was only seen in $p \bar{p} \to \omega\eta$ and $p \bar{p} \to \omega \pi^0 \pi^0$ \cite{Anisovich:2001cr,Anisovich:2002xoo}, while the $\rho_5(2350)$ has been identified in several reactions, including $\pi^{-} p \to \omega \pi^0 n$, $p\bar{p} \to \pi\pi$, and $p\bar{p} \to K^+ K^-$~\cite{GAMS:1994jop, Hasan:1994he, Carter:1978ux}.

{Several features can be inferred from the PDG values listed in Table~\ref{mass2}. For $J=2$, the isoscalar $\omega_2(1975)$ has a mass of $1975 \pm 20$ MeV and a width of $175 \pm 25$ MeV, while the isovector $\rho_2(1940)$ has a mass of $1940 \pm 40$ MeV and a width of $155 \pm 40$ MeV. The mass splitting between $\omega_2$ and $\rho_2$ is approximately $35$ MeV, which is relatively small, and their widths are comparable. 
Moving to $J=4$, the isoscalar $\omega_4(2250)$ and the isovector $\rho_4(2230)$ have comparable masses around $2.24\ \mathrm{GeV}$, with the $\rho_4$ being slightly lighter. Their total widths are of order $150$–$210\ \mathrm{MeV}$. 
For $J=5$, the $\omega_5(2250)$ and $\rho_5(2350)$ are heavier and exhibit significantly larger widths. Notably, the mass splitting between $\omega_J$ and $\rho_J$ partners is small for $J=4$, but becomes more pronounced for $J=5$, reaching about $80\ \mathrm{MeV}$, with the $\rho_5$ being the heavier state.}

\begin{table}[htbp]
\centering
\caption{{The experimental information of the $\omega_J/\rho_J$ ($J=2,4,5$) mesons.
Here, the masses and widths are average values
taken from the PDG \cite{ParticleDataGroup:2024cfk}.}  \label{mass2}}
\renewcommand\arraystretch{1.5}
\begin{tabular*}{80mm}{@{\extracolsep{\fill}}lccc}
\toprule[1.00pt]
\toprule[1.00pt]
 $ I(J^{PC})$ & States & Mass (MeV) & Width  (MeV)\\
\hline
   $0~(2^{--})$      & $\omega_2(1975)$ &$1975\pm20$&$175\pm25$ \\
   $1~(2^{--})$      & $\rho_2(1940)$   &$1940\pm40$&$155\pm40$ \\
   $0~(4^{--})$      & $\omega_4(2250)$ &$2250\pm30$&$150\pm50$ \\
   $1~(4^{--})$      & $\rho_4(2230)$   &$2230\pm25$&$210\pm30$ \\
   $0~(5^{--})$      & $\omega_5(2250)$ &$2250\pm70$&$320\pm95$ \\
   $1~(5^{--})$      & $\rho_5(2350)$   &$2330\pm35$&$400\pm100$ \\
\bottomrule[1pt]
\bottomrule[1pt]
\end{tabular*}
\end{table}

We now examine the theoretical decay properties of the $\omega_J$ and $\rho_J$ ($J=2,4,5$) mesons. Their predicted partial decay channels and corresponding branching ratios, calculated within the QPC model~\cite{Pang:2015eha}, are compiled in Table~\ref{decay2}.

{For the $\omega_2(1975)$, the dominant decay channel is $\pi\rho(1450)$ with a branching ratio of $0.48$, indicating a strong preference for decay into a pseudoscalar meson and an excited vector meson. The channel $\pi\rho$ also contributes significantly with a branching ratio of $0.24$. Other notable channels include $\pi b_1$ ($0.11$) and $\pi\rho_3(1690)$ ($0.07$). For the $\rho_2(1940)$, the largest decay channel is $\pi a_2(1320)$ with a branching ratio of $0.25$, highlighting decays into a pseudoscalar and a tensor meson. The channel $\pi\omega(1420)$ has a branching ratio of $0.21$, and the  $\pi\omega$ contributes $0.15$. Other significant channels include $\pi h_1(1170)$ and $\rho\eta$.}

For the $\omega_4(2250)$, the QPC model identifies the $a_1 \rho$ and $a_2 \rho$ channels as dominant, indicating a preference for decays into an axial-vector or tensor meson plus a vector meson. The decay $\omega f_2$ also contributes significantly. In contrast, the $\rho_4(2230)$ is predicted to decay predominantly into $\rho b_1$ and $\rho\rho$, highlighting decays involving an excited axial-vector $b_1$ or a pair of vector mesons.

For the $J=5$ states, the $\omega_5(2250)$ has a very large branching fraction into $\rho a_2$ (branching ratio $\simeq 0.58$), followed by $\omega f_2$. The $\rho_5(2350)$ does not exhibit a single overwhelmingly dominant channel; its largest decay channels are $\rho f_2$ and $\omega a_2$, each with a branching ratio of 0.24, alongside smaller but notable contributions from $\rho b_1$ and $\rho\rho$. These predictions provide a specific, testable framework for the decay properties of these higher-spin states.

 \begin{figure}[htbp]
    \includegraphics[width=1\linewidth]{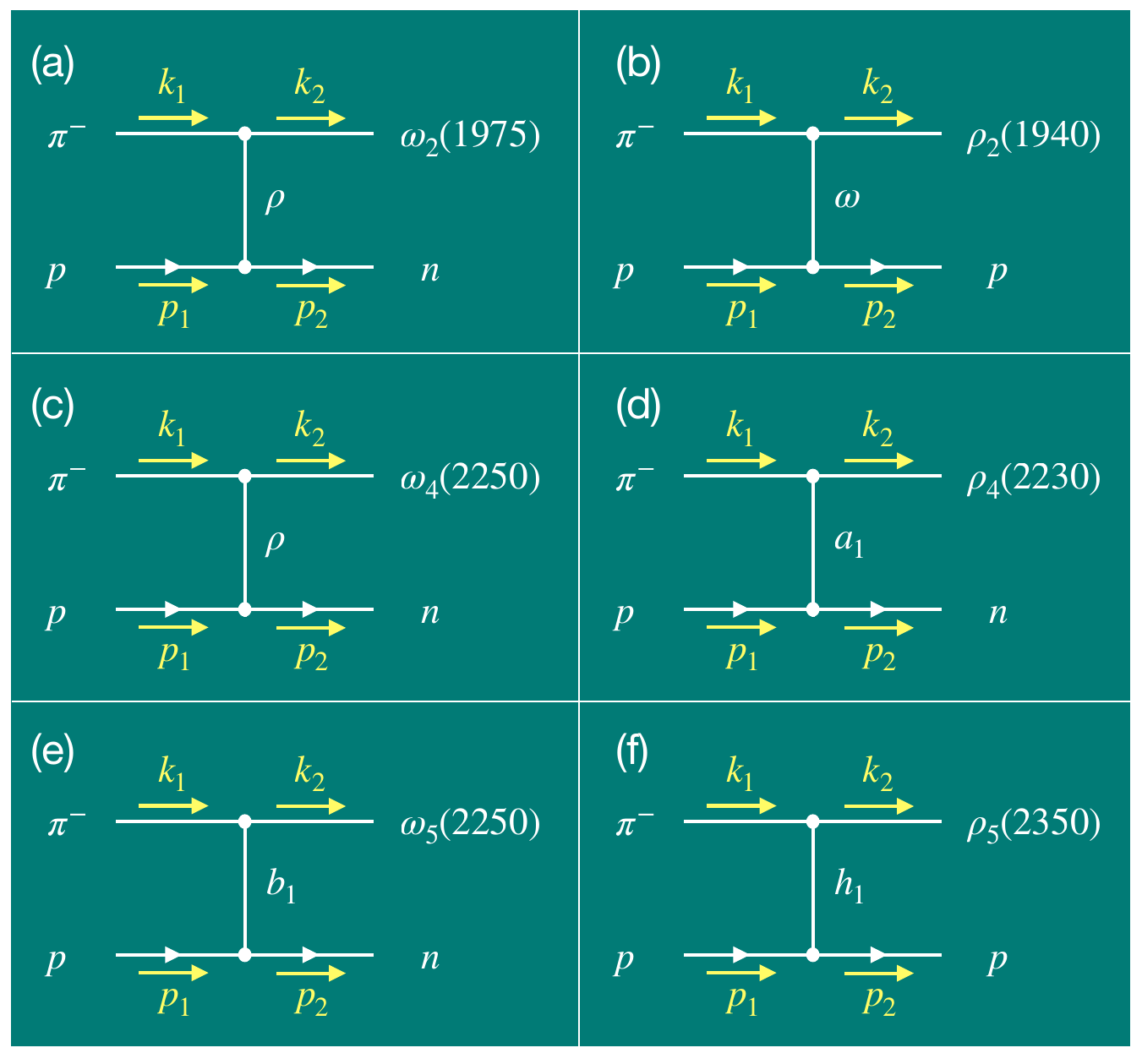}
\caption{ Feynman diagrams for the $\pi^- p \to \omega_J/\rho_J +N$ reactions.}
\label{Feynman diagrams}
\end{figure}

\medskip

\noindent
\textbf{Production mechanism.}
{Building on the successful description of $\omega_3(1670)$ and $\rho_3(1690)$ production in $\pi^- p$ scattering using a universal cutoff $\Lambda_t = 3.5 \pm 0.5\ \mathrm{GeV}$, the similiar framework is further employed to predict the production of the corresponding lower-spin and higher-spin partners.}

The phenomenological basis for this extension lies in the decay patterns predicted by the QPC model (Table~\ref{decay2}). Although not always the dominant modes, decays such as $\omega_4 \to \pi\rho$ and $\rho_4 \to \pi a_1$ occur with non-negligible branching ratios. {This indicates that these states can be produced in $\pi^- p$ reactions via $t$-channel single-meson exchange, in direct analogy to the $J=3$ case. The choice of the exchanged meson in the $t$-channel is guided by the dominant $\pi$-containing decay modes predicted by the QPC model (Table~\ref{decay2}). The corresponding tree-level Feynman diagrams for the reactions $\pi^- p \to \omega_2(1975) n$, $\pi^- p \to \rho_2(1940) p$, $\pi^- p \to \omega_4(2250) n$, $\pi^- p \to \rho_4(2230) n$, $\pi^- p \to \omega_5(2250) n$, and $\pi^- p \to \rho_5(2350)^-p$ are depicted in Fig.~\ref{Feynman diagrams}(a)-(f), respectively.}
 Given the limited existing experimental data on their production, these parameter-free predictions offer valuable guidance for future investigations in this energy region.

\begin{table*}[htbp]
\centering
\caption{The partial decay channels of the $\omega_J/\rho_J$ ($J=2,4,5$) mesons \cite{Pang:2015eha}. 
  \label{decay2}}
\renewcommand\arraystretch{1.5}
\begin{tabular*}{\textwidth}{@{\extracolsep{\fill}}lcccccccccccc}
\toprule[1.00pt]
\toprule[1.00pt]
 \multicolumn{2}{c}{$\omega_2(1975)$} & \multicolumn{2}{c}{$\rho_2(1940)$} & \multicolumn{2}{c}{$\omega_4(2250)$} & \multicolumn{2}{c}{$\rho_4(2230)$} & \multicolumn{2}{c}{$\omega_5(2250)$} & \multicolumn{2}{c}{$\rho_5(2350)$} \\
 \hline
 Channel &{\it BR} & Channel &{\it BR} & Channel & {\it BR} & Channel & {\it BR} & Channel & {\it BR} & Channel & {\it BR} \\
 \hline
$\pi\rho(1450)$ & 0.48 & $\pi a_2(1320)$ & 0.25 & $a_1\rho$ & 0.35 & $\rho b_1$ & 0.27 & $\rho a_2$ & 0.58 & $\rho f_2$ & 0.24 \\
$\pi\rho$ & 0.24 & $\pi\omega(1420)$ & 0.21 & $a_2\rho$ & 0.30 & $\rho\rho$ & 0.20 & $\omega f_2$ & 0.23 & $\omega a_2$ & 0.24 \\
$\pi b_1$ & 0.11 & $\pi\omega$ & 0.15 & $\omega f_2$ & 0.13 & $\rho f_2$ & 0.10 & $\pi b_1$ & 0.09 & $\rho b_1$ & 0.13 \\
$\pi\rho_3(1690)$ & 0.07 & $\pi h_1(1170)$ & 0.10 & $\pi\rho$ & 0.06 & $\pi a_1$ & 0.09 & $\pi a_1$ & 0.04 & $\rho\rho$ & 0.09 \\
$\omega\eta$ & 0.03 & $\rho\eta$ & 0.07 & $\pi\rho(1700)$ & 0.05 & $\omega a_1$ & 0.09 & $\rho\pi$ & 0.03 & $\pi\pi_2$ & 0.07 \\
$\pi\rho(1700)$ & 0.02 & $\pi a_0(1450)$ & 0.06 & $\omega f_1$ & 0.04 & $\omega a_2$ & 0.06 & $\eta h_1$ & 0.01 & $\pi a_2$ & 0.07 \\
$KK^*$ & 0.02 & $\rho\rho$ & 0.05 & $\rho\rho$ & 0.03 & $\pi a_2$ & 0.04 & & & $\pi h_1$ & 0.05 \\
$KK^*(1410)$ & 0.02 & $KK^*$ & 0.04 & & & $\pi\omega$ & 0.01 & & & $\pi a_1$ & 0.04 \\
 & & $KK^*_2(1430)$ & 0.03 & & & & & & & $\pi\omega$ & 0.01 \\
 & & $\rho\eta^{\prime}$ & 0.03 & & & & & & & & \\
\bottomrule[1pt]
\bottomrule[1pt]
\end{tabular*}
\end{table*}
All upper vertices (three-meson couplings) in Fig.~\ref{Feynman diagrams} are constructed in this work. {Their effective Lagrangians, written in terms of the tensor fields $W_J$ for the high-spin states, are given below.} The lower vertices (meson-nucleon-nucleon couplings) are taken from the established phenomenological Lagrangians~\cite{Rijken:2010zzb}.

\paragraph*{$\pi^{-} p \to \omega_2(1975) n$ }[Fig.~\ref{Feynman diagrams}(a)].
{This reaction proceeds via $\rho$-meson exchange, motivated by the fact that $\pi\rho$ is the most prominent $\pi$-containing decay of $\omega_2(1975)$ (Table~\ref{decay2}). The effective Lagrangian for the $W_2V_1P$ vertex is
\begin{equation}
\mathcal{L}_{W_{2} V_{1} P}=g_{W_{2} V_{1} P} \operatorname{tr}\left[W_{2}^{\alpha \beta} \left\{ V_{\alpha}, \partial_{\beta} P\right\}_{+}\right],
\label{eq:omega2}
\end{equation}
where the fields $W_2$, $V_1$, and $P$ correspond to the spin-$2$ meson with $J^{PC}=2^{--}$, vector meson and pseudoscalar meson, respectively. The nonet of $2^{--}$ reads
$$
W_{2}^{\mu\nu}=\frac{1}{\sqrt{2}}\left(\begin{array}{ccc}
\frac{\omega_{2}^{\mu\nu}+\rho_{2}^{0\mu\nu}}{\sqrt{2}} &\rho_{2}^{+\mu\nu} &K_{2}^{+\mu\nu}\\
\rho_{2}^{-\mu\nu} &\frac{\omega_{2}^{\mu\nu}-\rho_{2}^{0\mu\nu}}{\sqrt{2}} &K_{2}^{0\mu\nu}\\
K_{2}^{-\mu\nu} &\bar{K}_{2}^{0\mu\nu} &\omega_{2}^{\mu\nu}
\end{array}\right).
$$
The specific form  of the $\omega_2 \rho \pi $ vertex is as follows:
\begin{equation}
\mathcal{L}_{\omega_{2} \rho \pi}=\frac{1}{2} g_{\omega_{2} \rho \pi} \omega_{2}^{\alpha \beta} \rho_{\alpha}^{+} \partial_{\beta} \pi^{-}.
\label{eq:omega21}
\end{equation}
The $\rho NN$ vertex is presented in Eq.~(\ref{eq:rhoNN}).}

\paragraph*{$\pi^{-} p \to \rho_2(1940) p$} [Fig.~\ref{Feynman diagrams}(b)].
{The forms of the Lagrangian for the two vertices in this process are exactly the same as that in the previous process ($\pi^{-} p \to \omega_2(1975) n$), with only the exchanged particle changed from $\rho$ to $\omega$. Hence, we omit the Lagrangian required for this Feynman process here.}

\paragraph*{$\pi^{-} p \to \omega_4(2250) n$ }[Fig.~\ref{Feynman diagrams}(c)].
This reaction proceeds via $\rho$-meson exchange, motivated by the fact that $\pi\rho$ is the most prominent $\pi$-containing decay of $\omega_4(2250)$ (Table~\ref{decay2}). The effective Lagrangian for the $W_4V_1P$ vertex is
\begin{equation}
\mathcal{L}_{W_4V_1 P}=g_{W_4V_1 P} \operatorname{tr}\!\left[ W_4^{\mu\nu\alpha\beta} \left\{ V_\mu, \partial_\nu\partial_\alpha\partial_\beta P \right\}_+ \right],
\label{eq:omega4}
\end{equation}
where the fields $W_4$, $V_1$, and $P$ correspond to the spin-$4$ meson with $J^{PC}=4^{--}$, vector meson and pseudoscalar meson, respectively. The nonet of $4^{--}$ reads
$$
W_{4}^{\mu\nu\rho\sigma}=\frac{1}{\sqrt{2}}\left(\begin{array}{ccc}
\frac{\omega_{4}^{\mu\nu\rho\sigma}+\rho_{4}^{0\mu\nu\rho\sigma}}{\sqrt{2}} &\rho_{4}^{+\mu\nu\rho\sigma} &K_{4}^{+\mu\nu\rho\sigma}\\
\rho_{4}^{-\mu\nu\rho\sigma} &\frac{\omega_{4}^{\mu\nu\rho\sigma}-\rho_{4}^{0\mu\nu\rho\sigma}}{\sqrt{2}} &K_{4}^{0\mu\nu\rho\sigma}\\
K_{4}^{-\mu\nu\rho\sigma} &\bar{K}_{4}^{0\mu\nu\rho\sigma} &\omega_{4}^{\mu\nu\rho\sigma}
\end{array}\right).
$$
The specific form  of the $\omega_4 \rho \pi $ vertex is as follows:
\begin{equation}
\mathcal{L}_{\omega_4 \rho \pi} = \frac{1}{2} \, g_{\omega_4 \rho \pi} \, \omega_4^{\sigma\alpha\beta\gamma} \, \rho^+_{\sigma} \, \partial_\alpha \partial_\beta \partial_\gamma \pi^-.
\label{eq:omega41}
\end{equation}
The $\rho NN$ vertex is presented in Eq.~(\ref{eq:rhoNN}).

\paragraph*{$\pi^{-} p \to \rho_4(2230) n$ }[Fig.~\ref{Feynman diagrams}(d)].
The dominant $\pi$-containing decay $\pi a_1$ leads to $a_1$-meson exchange. The effective Lagrangian is
\begin{equation}
\mathcal{L}_{W_4 A_1 P}=g_{W_4 A_1 P} \, \varepsilon_{\mu\nu\rho\sigma}\operatorname{tr}\!\Big[W_4^{\mu\alpha\beta\gamma}\big[(\partial^{\nu} A_1^{\rho}), (\partial_{\alpha}\partial_{\beta}\partial_{\gamma}\partial^{\sigma} P)\big]_{-}\Big],
\label{eq:rho4a1pi}
\end{equation}
where the fields $W_4$, $A_1$, and $P$ correspond to the spin-$4$ meson with $J^{PC}=4^{--}$, axial-vector meson and pseudoscalar meson, respectively. The matrix $A_1$ has the form
$$ A_{1}^{\mu}=\frac{1}{\sqrt{2}}\left(\begin{array}{ccc}
\frac{f_{1}^{\mu}+a_{1}^{0\mu}}{\sqrt{2}} &a_{1}^{+\mu} &K_{1,A}^{+\mu}\\
a_{1}^{-\mu} &\frac{f_{1}^{\mu}-a_{1}^{0\mu}}{\sqrt{2}} &K_{1,A}^{0\mu}\\
K_{1, A}^{-\mu} &\bar{K}_{1, A}^{0\mu} &f_{1}^{\mu}
\end{array}\right). $$
The Lagrangian for $\rho_4 a_1 \pi$ vertex can be expressed in the following specific form:
\begin{equation}
\mathcal{L}_{\rho_4 a_1 \pi} = \frac{1}{2} \, g_{\rho_4 a_1 \pi} \, \epsilon_{\mu\nu\rho\sigma} \, \rho_4^{\mu\alpha\beta\gamma} \, (\partial^\nu a_1^{\rho+}) \, \partial_\alpha \partial_\beta \partial_\gamma \partial^\sigma \pi^- ,
\label{eq:rho4a1pi1}
\end{equation}
and the $a_1 NN$ vertex is taken from Ref.~\cite{Rijken:2010zzb}:
\begin{equation}
\mathcal{L}_{a_1 N N}=g_{a_1N N} \big[\bar{N} \gamma_{\mu} \gamma_5 N\big]a_1^{\mu} + \frac{i f_{a_1N N}}{m_P} \big[\bar{N} \gamma_5 N\big] \partial_{\mu}a_1^{\mu}.
\label{eq:a1NN}
\end{equation}

\paragraph*{$\pi^{-} p \to \omega_5(2250) n$ }[Fig.~\ref{Feynman diagrams}(e)].
This channel is dominated by $t$-channel $b_1$ exchange. The corresponding Lagrangian is
\begin{equation}
\mathcal{L}_{W_5 B_1 P} = g_{W_5 B_1 P} \, \operatorname{tr}\!\Big[ W_5^{\mu\alpha\beta\gamma\delta} \big\{ B_{1,\mu},\; \partial_\alpha\partial_\beta\partial_\gamma\partial_\delta P \big\}_+ \Big].
\label{eq:omega5b1pi}
\end{equation}
where the fields $W_5$, $B_1$, and $P$ correspond to the spin-$5$ meson with $J^{PC}=5^{--}$, pseudovector meson and pseudoscalar meson, respectively.
The matrix $W_5$ and $B_1$ have the form
$$
W_{5}^{\mu\nu\rho\sigma\gamma}=\frac{1}{\sqrt{2}}\left(\begin{array}{ccc}
\frac{\omega_{5N}^{\mu\nu\rho\sigma\gamma}+\rho_{5}^{0\mu\nu\rho\sigma\gamma}}{\sqrt{2}} &\rho_{5}^{+\mu\nu\rho\sigma\gamma} &K_{5}^{+\mu\nu\rho\sigma\gamma}\\
\rho_{5}^{-\mu\nu\rho\sigma\gamma} &\frac{\omega_{5}^{\mu\nu\rho\sigma\gamma}-\rho_{5}^{0\mu\nu\rho\sigma\gamma}}{\sqrt{2}} &K_{5}^{0\mu\nu\rho\sigma\gamma}\\
K_{5}^{-\mu\nu\rho\sigma\gamma} &\bar{K}_{5}^{0\mu\nu\rho\sigma\gamma} &\omega_{5}^{\mu\nu\rho\sigma\gamma}
\end{array}\right),
$$
$$ B_{1}^{\mu}=\frac{1}{\sqrt{2}}\left(\begin{array}{ccc}
\frac{h_{1}^{\mu}+b_{1}^{0\mu}}{\sqrt{2}} &b_{1}^{+\mu} &K_{1,B}^{+\mu}\\
b_{1}^{-\mu} &\frac{h_{1}^{\mu}-b_{1}^{0\mu}}{\sqrt{2}} &K_{1,B}^{0\mu}\\
K_{1, B}^{-\mu} &\bar{K}_{1, B}^{0\mu} &h_{1}^{\mu}
\end{array}\right). $$
The Lagrangian for $\omega_5 b_1 \pi$ vertex can be expressed in the following specific form:
\begin{equation}
\mathcal{L}_{\omega_5 b_1 \pi} = \frac{1}{2} \, g_{\omega_5 b_1 \pi} \, \omega_5^{\sigma\alpha\beta\gamma\delta} \, b_{1,\sigma}^{+} \, \partial_\alpha \partial_\beta \partial_\gamma \partial_\delta \pi^- .
\label{eq:omega5b1pi1}
\end{equation}
The $b_1 NN$ vertex is adopted from Ref.~\cite{Rijken:2010zzb}:
\begin{equation}
\mathcal{L}_{b_1 N N} = \frac{i f_{b_1 N N}}{m_{b_1}} \left[ \bar{N} \sigma_{\mu\nu} \gamma_5 N \right] \partial^{\nu} b_1^{\mu}.
\label{eq:b1NN}
\end{equation}

\paragraph*{$\pi^{-} p \to \rho_5(2350) p$} [Fig.~\ref{Feynman diagrams}(f)].
Here the exchange of an $h_1$ meson is involved, following the QPC prediction that $\pi h_1$ is the most important $\pi$-containing channel. The effective Lagrangian is
\begin{equation}
\mathcal{L}_{W_5 B_1 P} = g_{W_5 B_1 P} \, \operatorname{tr}\!\Big[ W_5^{\mu\alpha\beta\gamma\delta} \big\{ B_{1,\mu},\; \partial_\alpha\partial_\beta\partial_\gamma\partial_\delta P \big\}_+ \Big].
\label{eq:rho5h1pi}
\end{equation}
The Lagrangian for $\rho_5 h_1 \pi$ vertex can be expressed in the following specific form:
\begin{equation}
\mathcal{L}_{\rho_5 h_1 \pi} = \frac{1}{2} \, g_{\rho_5 h_1 \pi} \, \rho_5^{\mu\alpha\beta\gamma\delta} \, h_{1,\mu} \, \partial_\alpha \partial_\beta \partial_\gamma \partial_\delta \pi^- .
\label{eq:rho5h1pi1}
\end{equation}
The $h_1 NN$ vertex, which has the same tensor structure as Eq.~(\ref{eq:b1NN}), is also taken from Ref.~\cite{Rijken:2010zzb}.

All coupling constants for the vertices appearing in the above Lagrangians are listed in Table~\ref{cp}.

\begin{table}[htbp]
\centering
\caption{Coupling constants used in this work. The values in the left column are the three-meson couplings determined in this work by fitting the decay widths from Table~\ref{decay2}. Those in the right column are the meson-nucleon-nucleon couplings taken from Ref.~\cite{Rijken:2010zzb}. They correspond to the upper and lower vertices, respectively, in the Feynman diagrams of Fig.~\ref{Feynman diagrams}. \label{cp}}
\renewcommand\arraystretch{1.8}
\setlength{\tabcolsep}{15pt}
\begin{tabular}{lr}
\toprule[1.00pt]
\toprule[1.00pt]
$g_{\rho_2\omega\pi}=5.89$& $g_{\omega NN}=11.04$\\ 
&$f_{\omega NN}=-2.02$\\
$g_{\omega_2\rho\pi}=4.39$& $g_{\rho NN}=2.05$\\ 
$g_{\omega_4\rho\pi}=3.91\ \mathrm{GeV}^{-2}$&$f_{\rho NN}=13.40$\\
  $g_{\rho_4 a_1 \pi}=18.82\ \mathrm{GeV}^{-4}$&$g_{a_1N N}=-2.90$ \\
 &$f_{a_1N N}=-5.86$\\
$g_{\omega_5 b_1 \pi} = 45.16\ \mathrm{GeV}^{-3}$  &$f_{b_1 N N}=-8.01$ \\
$g_{\rho_5 h_1 \pi}=55.27\ \mathrm{GeV}^{-3}$ &$f_{h_1 NN}=-0.29$ \\
\bottomrule[1pt]
\bottomrule[1pt]
\end{tabular}
\end{table}

{Using the effective Lagrangians introduced in the previous section, we write the tree-level amplitudes for the six reactions as follows.}

{
\paragraph*{$\pi^- p \to \omega_2(1975) n$ via $t$-channel $\rho$ exchange:}
\begin{equation}
\begin{split}
i\mathcal{M}_{\omega_2} = & -\frac{i}{2} g_{\omega_{2} \rho \pi} \varepsilon^{\alpha \beta *}(k_{2}) ( k_{1 \beta}) \frac{-i(g^{\mu}_{ \alpha} + q_{\mu} q_{\alpha} / m_{\rho}^{2})}{t - m_{\rho}^{2}}\\
&\times \bar{u}(p_{2})\left( g_{\rho NN} \gamma_{\mu} + \frac{f_{\rho NN}}{2 m_{p}} \gamma_{\mu} \slashed{q} \right) u(p_{1}) F_{t}(q),
\end{split}
\label{Momega2}
\end{equation}}

{\paragraph*{$\pi^- p \to \rho_2(1940) p$ via $t$-channel $\omega$ exchange:}
\begin{equation}
\begin{split}
i\mathcal{M}_{\rho_2} = & -\frac{i}{2} g_{\rho_{2} \omega \pi} \varepsilon^{\alpha \beta *}(k_{2}) ( k_{1 \beta}) \frac{-i(g^{\mu}_{ \alpha} + q_{\mu} q_{\alpha} / m_{\omega}^{2})}{t - m_{\omega}^{2}}\\
&\times \bar{u}(p_{2})\left( g_{\omega NN} \gamma_{\mu} + \frac{f_{\omega NN}}{2 m_{p}} \gamma_{\mu} \slashed{q} \right) u(p_{1}) F_{t}(q),
\end{split}
\label{Mrho2}
\end{equation}}

\paragraph*{$\pi^- p \to \omega_4(2250) n$ via $t$-channel $\rho$ exchange:}
\begin{equation}
\begin{split}
i\mathcal{M}_{\omega_4} = & \frac{i}{2}\,g_{\omega_4 \rho \pi} \, k_{1\alpha} k_{1\beta} k_{1\gamma}\,\varepsilon^{\sigma\alpha\beta\gamma*}(k_{2}) 
\frac{-i \left( g^{\mu}_{\sigma} + q^{\mu} q_{\sigma} / m_{\rho}^{2}\right)}{t-{m_{\rho}}^{2}} \\
&\times\bar{u}(p_{2}) \left[ g_{\rho NN}\gamma_{\mu} +\frac{f_{\rho NN}}{2m_{p}} \gamma_{\mu} \slashed{q}\right] u(p_{1}) \, F_t(q),
\end{split}
\label{Momega4}
\end{equation}

\paragraph*{$\pi^- p \to \rho_4(2230) n$ via $t$-channel $a_1$ exchange:}
\begin{equation}
\begin{split}
i\mathcal{M}_{\rho_4} = & -\frac{1}{2} \, g_{ \rho_4 a_{1} \pi} \, \epsilon_{\mu\nu\rho\sigma}  q^{\nu}k_{1\alpha} k_{1\beta} k_{1\gamma} k_1^{\sigma} 
 \\
&\varepsilon^{\mu\alpha\beta\gamma*}(k_2)\frac{-i \left( g^{\xi}_{\rho} + q^{\xi} q_{\rho} / m_{a_1}^{2}\right)}{t - m_{a_1}^2} \, F_t(q)\\
&\times\bar{u}(p_2) \left( g_{a_1 NN} \gamma_{\xi}\gamma_5 - \frac{f_{a_1 NN}}{M_p} \gamma_5 \slashed{q} \right) u(p_1),
\end{split}
\label{Mrho4}
\end{equation}

\paragraph*{$\pi^- p \to \omega_5(2250) n$ via $t$-channel $b_1$ exchange:}
\begin{equation}
\begin{split}
i\mathcal{M}_{\omega_5} = &\frac{1}{2}\,\frac{f_{b_1 N N}}{m_{b_{1}}}\, g_{\omega_5 b_{1} \pi}  \, k_{1\alpha} k_{1\beta} k_{1\gamma} k_{1\delta} \, \varepsilon^{\sigma\alpha\beta\gamma\delta*}(k_{2}) \, \\
&\times\frac{-i \left( g^{\mu}_{\sigma} + q^{\mu} q_{\sigma} / m_{b_1}^{2}\right)}{t-m_{b_{1}}^{2}} \,
\bar{u}(p_{2}) \sigma_{\mu\nu} \gamma_{5} u(p_{1}) \\
&\times q^{\nu} F_t(q),
\end{split}
\label{Momega5}
\end{equation}

\paragraph*{$\pi^- p \to \rho_5(2350) n$ via $t$-channel $h_1$ exchange:}
\begin{equation}
\begin{split}
i\mathcal{M}_{\rho_5} = &-\frac{1}{2}\,\frac{f_{h_1 N N}}{m_{b_{1}}}\, g_{\rho_5  h_{1} \pi}  \, k_{1\alpha} k_{1\beta} k_{1\gamma} k_{1\delta} \, \varepsilon^{\sigma\alpha\beta\gamma\delta*}(k_{2}) \, \\
&\times\frac{-i \left( g^{\mu}_{\sigma} + q^{\mu} q_{\sigma} / m_{h_1}^{2}\right)}{t-m_{h_{1}}^{2}} \,
\bar{u}(p_{2}) \sigma_{\mu\nu} \gamma_{5} u(p_{1}) \\
&\times q^{\nu} F_t(q).
\end{split}
\label{Mrho5}
\end{equation}

\begin{figure*}[htbp]
\includegraphics[width=1\linewidth]{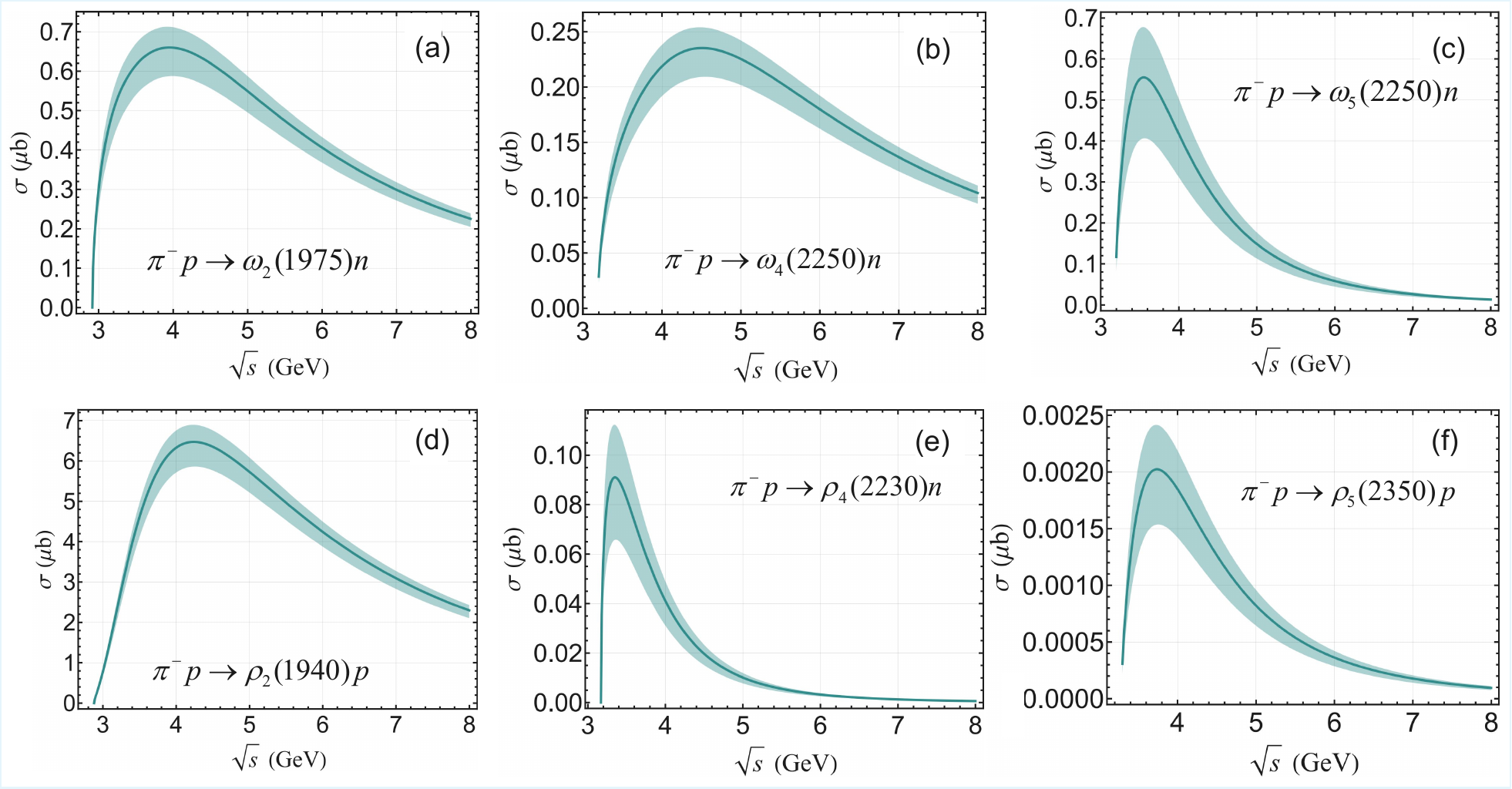} 
\caption{{The energy dependence of the total cross
section for production of the $\omega_2(1975)$,  $\rho_2(1940)$,  $\omega_4(2250)$,  $\rho_4(2230)$, $\omega_5(2250)$,  $\rho_5(2350)$ through $t$-channel with cutoff
$\Lambda_t =3.5\pm0.5$ GeV.}}
\label{higher}
\end{figure*}
In the above expressions, $\varepsilon(k_2)$ denotes the polarization tensor of the outgoing high-spin meson. $\bar{u}(p_2)$ and $u(p_1)$ are the Dirac spinors of the outgoing and incoming nucleons, respectively. $t=q^2=(k_1-k_2)^2$ is the squared four-momentum transfer. The form factor $F_t(q)$ is taken from Eq.~(\ref{Ft}) with the same cutoff $\Lambda_t=3.5\pm0.5\;\mathrm{GeV}$ used for the $J=3$ states.

{The $t$-channel propagators are Reggeized according to Eq.~(\ref{rho_propagator}). The corresponding Regge trajectories for the exchanged mesons are
\begin{equation}
\alpha_R(t) = 1 + \alpha'_R\left(t - m_R^2\right), \qquad R = \rho,\;\omega,\;a_1,\;b_1,\;h_1,
\label{alphaR}
\end{equation}
with slopes $\alpha'_\rho=0.8\;\mathrm{GeV}^{-2}$ \cite{Guidal:1997hy}, $\alpha'_\omega=0.9\;\mathrm{GeV}^{-2}$ \cite{Guidal:1997hy}, $\alpha'_{a_1}=0.8\;\mathrm{GeV}^{-2}$ (determined from the mass of the $a_3(1875)$ state following Ref.~\cite{Galata:2011bi}), $\alpha'_{b_1}=0.7\;\mathrm{GeV}^{-2}$ \cite{Wang:2025rvr}, and $\alpha'_{h_1}=0.7\;\mathrm{GeV}^{-2}$ \cite{Wang:2025rvr}.}

{With the amplitudes~(Eqs. (\ref{Momega4})--(\ref{Mrho5})) and the differential cross-section formula in Eq.~(\ref{diff}), we can now compute the total and differential cross sections for the six states. }

\begin{figure*}[htbp]
\includegraphics[width=1\linewidth]{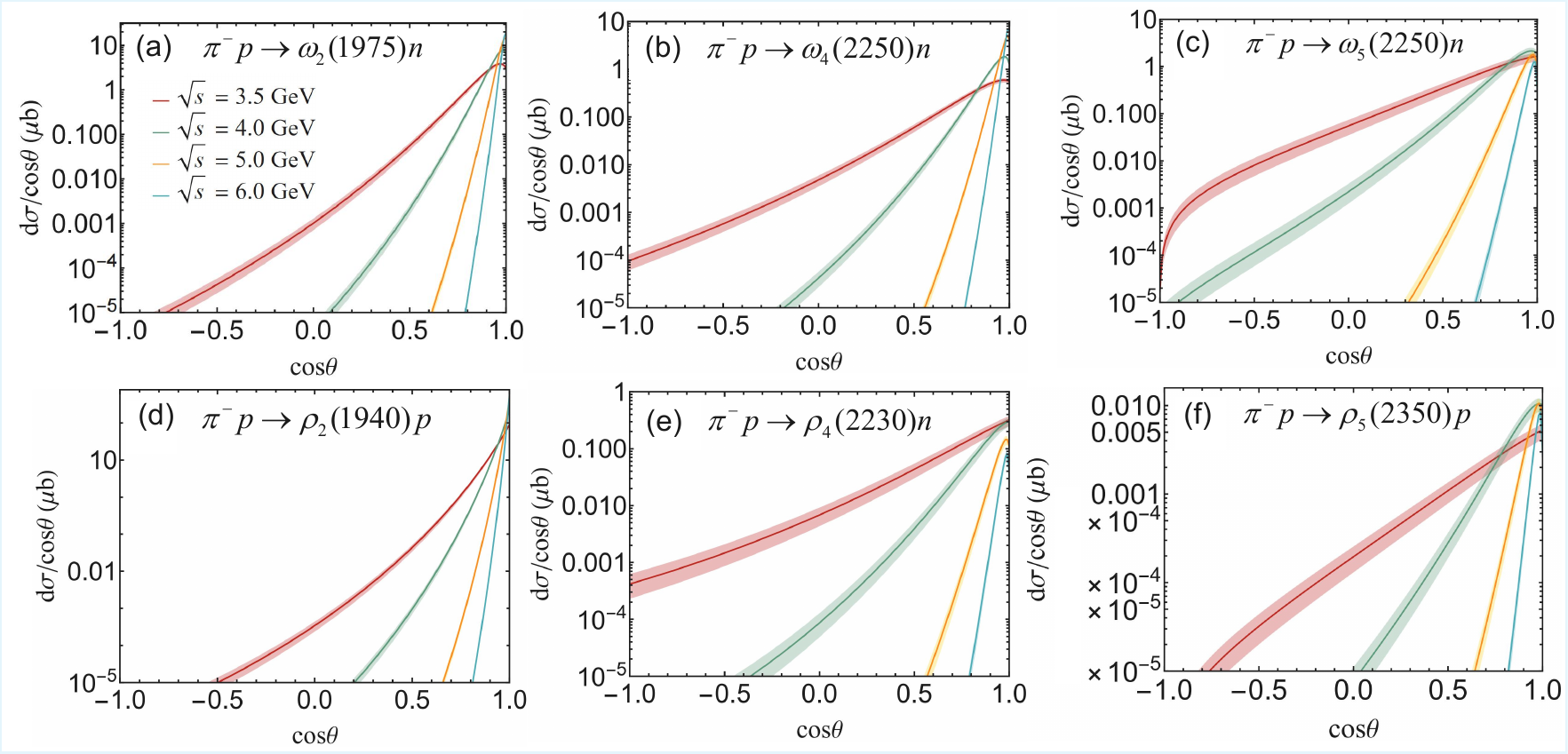} 
\caption{{The differential cross section $d\sigma/d\cos\theta$
of the  $\omega_2(1975)$,  $\rho_2(1940)$, $\omega_4(2250)$, $\rho_4(2230)$, $\omega_5(2250)$,   $\rho_5(2350)$ production at different c.m. energies W =
3.5, 4, 5 and 6 GeV.}}
\label{highercs}
\end{figure*}

\textbf{Total cross sections.}
{The total cross sections for meson--nucleon production of the $J=2$ states $\omega_2(1975)$ and $\rho_2(1940)$, the $J=4$ states $\omega_4(2250)$ and $\rho_4(2230)$, and the $J=5$ states $\omega_5(2250)$ and $\rho_5(2350)$ via $t$-channel exchange are shown in Fig.~\ref{higher} as functions of the center-of-mass energy $\sqrt{s}$.} The solid curves correspond to a central cutoff $\Lambda_t = 3.5~\mathrm{GeV}$, with the shaded bands representing the uncertainty from varying $\Lambda_t$ by $\pm 0.5~\mathrm{GeV}$.

{
For the \(J=2\) states, subplots (a) and (d) illustrate clear spin-isospin patterns. The \(\omega_2(1975)\) cross section  peaks at approximately \(0.7~\mu\mathrm{b}\) near \(\sqrt{s} \approx 3.6~\mathrm{GeV}\), while the \(\rho_2(1940)\) cross section exhibits a significantly larger peak of about \(6.5~\mu\mathrm{b}\) at a similar energy. This order-of-magnitude difference in peak cross sections for states with identical spin \(J=2\) but different isospin underscores a strong isospin dependence in the production mechanisms. Both curves show a broad peak structure followed by a monotonic decline at high \(\sqrt{s}\). 
A clear spin-dependent and isospin-dependent pattern is observed. The $J=4$ cross sections [subplots (b) and (e)] are of the order of $0.1$--$0.3~\mu\mathrm{b}$.} $\omega_4(2250)$ exhibits a broad peak of about $0.25~\mu\mathrm{b}$ near $\sqrt{s}=4.3~\mathrm{GeV}$, while $\rho_4(2230)$ peaks at a lower value of approximately $0.14~\mu\mathrm{b}$. {In contrast, the $J=5$ cross sections [subplots (c) and (f)] display a significant isospin splitting.} The $\omega_5(2250)$ cross section reaches a maximum of about $0.68~\mu\mathrm{b}$, comparable to the $J=4$ cross sections, whereas $\rho_5(2350)$ is strongly suppressed, with a peak of only about $0.0025~\mu\mathrm{b}$. The suppression of the $\rho_5$ state, despite its higher mass, is a notable feature that may be attributed to the specific coupling structure and the higher derivative nature of the interaction vertices for this channel. All curves decline monotonically at high $\sqrt{s}$, consistent with $t$-channel exchange kinematics. The narrow uncertainty bands indicate that the predictions are robust under variations of the cutoff parameter.

\textbf{Differential cross sections.}
The angular distributions are presented in Fig.~\ref{highercs}, which shows $d\sigma/d\cos\theta$ for the six states at center-of-mass energies $W = 3.5$, $4.0$, $5.0$, and $6.0\,\mathrm{GeV}$.

All panels in Fig.~\ref{highercs} share two key features of $t$-channel processes: (i) a very strong forward peak ($\cos\theta \to 1$) at all energies, and (ii) the forward peak narrows with increasing energy. This reflects the kinematics of meson exchange, where small momentum transfer favors scattering at very small angles.

A closer inspection reveals significant effects of both spin and isospin.
For the $J=2$ resonances, a clear isospin splitting is observed. The $\omega_2(1975)$ [Fig.~\ref{highercs}(a)] exhibits forward differential cross sections on the order of $10\,\mu\mathrm{b}$ across the measured energy range. In contrast, its isospin partner, the $\rho_2(1940)$ [Fig.~\ref{highercs}(d)], displays a consistently larger magnitude, with forward cross sections on the order of $10^{2}\,\mu\mathrm{b}$. This indicates that the production of the isovector $\rho_2$ state is significantly enhanced compared to the isoscalar $\omega_2$ state for $J=2$.

The $J=4$ state $\omega_4(2250)$ [Fig.~\ref{highercs}(b)] reaches differential cross sections of up to about $10;\mu\mathrm{b}$ in the extreme forward direction, while its isospin partner $\rho_4(2230)$ [Fig.~\ref{highercs}(e)] shows a similar shape but with a slightly gentler slope and a lower peak magnitude. The $J=5$ states [Figs.~\ref{highercs}(c) and (f)] exhibit the expected strong suppression, particularly for the $\rho_5(2350)$ state.

\section{Summary}
\label{section4}

In this work, we have performed a systematic study of the production of high-spin $\omega_J$ and $\rho_J$ mesons ($J =2, 3,4,5$) in $\pi^- p$ scattering, employing an effective Lagrangian approach. Our primary goal was to provide theoretical predictions for the production dynamics of these states, which are far less explored than their decay properties.

Our analysis is based on a unified framework that considers only the dominant $t$-channel meson-exchange mechanisms. {For each channel, the exchanged meson (e.g., $\rho$, $\pi$, $\omega$, $a_1$, $b_1$, or $h_1$) is selected according to the $\pi$-containing decay modes predicted by the QPC model. For the well-established $J=3$ states $\omega_3(1670)$ and $\rho_3(1690)$, the model has a single adjustable parameter: the form-factor cutoff $\Lambda_t$. A fit to the available total and differential cross-section data yields $\Lambda_t = 3.5 \pm 0.5\ \text{GeV}$. Using this determined parameter, we subsequently made parameter-free predictions for the production of their lower-spin and higher-spin partners $\omega_2(1975)$, $\rho_2(1940)$, $\omega_4(2250)$, $\rho_4(2230)$, $\omega_5(2250)$, and $\rho_5(2350)$.} The high-energy behavior is consistently described by applying Reggeization to the $t$-channel propagators.

The key findings of this study are as follows:
\begin{enumerate}
    \item The model with $\Lambda_t = 3.5 \pm 0.5\ \text{GeV}$ provides an excellent description of the existing total and differential cross-section data for $\pi^- p \to \omega_3(1670)n$ and $\pi^- p \to \rho_3(1690)n$, validating the theoretical framework.
    \item The predicted cross sections for the higher-spin ($J=4,5$) states are systematically suppressed compared to the $J=3$ case, an expected consequence of the higher derivative couplings in the production vertices. Crucially, all predicted cross sections remain in the experimentally accessible range.
    \item The differential cross sections for all states exhibit a pronounced forward peak ($\cos\theta \to 1$), a distinctive signature of the $t$-channel production mechanism.
    \item The consistent description of {eight} different resonance production channels with a single universal parameter demonstrates the robustness of the model. It provides a coherent picture linking the decay properties (from the QPC model) to the production dynamics in scattering experiments.
\end{enumerate}

In conclusion, this work provides the first comprehensive set of theoretical predictions for the production cross sections of the high-spin $\omega_2$, $\rho_2$, $\omega_4$, $\rho_4$, $\omega_5$, and $\rho_5$ states in $\pi^- p$ reactions. The results offer concrete, testable guidance for future experimental searches at facilities such as J-PARC and COMPASS, particularly in the forward-angle region where the cross sections are largest. This analysis bridges spectroscopy and production studies, supporting future meson-beam experiments in completing the spectrum of light hadrons.

\begin{acknowledgments}

T.Y.L. would like to thank Yu Zhuge, Shahriyar Jafarzade, and Xiao-Yun Wang for their useful discussions. This work is also supported by the National Natural Science Foundation of China under Grant Nos. 12335001 and 12247101, the ‘111 Center’ under Grant No. B20063, the Natural Science Foundation of Gansu Province (No. 22JR5RA389, No. 25JRRA799), the fundamental Research Funds for the Central Universities, the project for top-notch innovative talents of Gansu province, and Lanzhou City High-Level Talent Funding.

\end{acknowledgments}

\vfil

\section*{APPENDIX: EXPLICIT EXPRESSIONS OF PROPAGATORS}\label{APPENDIX B}

The propagator of a field for a boson field can be written as 
\begin{align}
S(p) = \frac{1}{p^2 - M^2} (\slashed{p} + M) 
\Delta_{\alpha_1^{} \cdots \alpha_{n-1}^{}}^{\beta_1^{} \cdots \beta_{n-1}^{}}(j,p).
\end{align}
The explicit expressions for the projection operators of fields from spin$-3$ to spin$-5$ are given as follows {\cite{Doring:2025sgb,Kim:2024mqx}}.
\begin{widetext}
\begin{align}
\Delta_{\alpha_{1}\alpha_{2}}^{\beta_{1}\beta_{2}}(2, p)=\frac{1}{2}\left(\bar{g}_{\alpha_{1}}^{\beta_{1}}\bar{g}_{\alpha_{1}}^{\beta_{2}}+\bar{g}_{\alpha_{1}}^{\beta_{1}}\bar{g}_{\alpha_{2}}^{\beta_{1}}-\frac{2}{3}\bar{g}_{\alpha_{1}\alpha_{2}}\bar{g}^{\beta_{1}\beta_{2}}\right).
\end{align}
\begin{align}
\Delta_{\alpha_1^{} \alpha_2^{} \alpha_3^{}}^{\beta_1^{} \beta_2^{} \beta_3^{}} (3,p) =& 
- \frac{1}{36} \sum_{P(\alpha), P(\beta)}
\left(
\bar{g}_{\alpha_1^{}}^{\beta_1^{}} \bar{g}_{\alpha_2^{}}^{\beta_2^{}}
\bar{g}_{\alpha_3^{}}^{\beta_3^{}}
- \frac35 \bar{g}_{\alpha_1^{} \alpha_2^{}}^{} \bar{g}^{\beta_1^{}\beta_2^{}}
\bar{g}_{\alpha_3^{}}^{\beta_3^{}}
\right)
\cr
=& -\frac{1}{6} \Big(
  \bar{g}_{\alpha_1}^{\beta_1} \bar{g}_{\alpha_2}^{\beta_2} \bar{g}_{\alpha_3}^{\beta_3}
+ \bar{g}_{\alpha_1}^{\beta_1} \bar{g}_{\alpha_2}^{\beta_3} \bar{g}_{\alpha_3}^{\beta_2}
+ \bar{g}_{\alpha_1}^{\beta_2} \bar{g}_{\alpha_2}^{\beta_1} \bar{g}_{\alpha_3}^{\beta_3}
+ \bar{g}_{\alpha_1}^{\beta_2} \bar{g}_{\alpha_2}^{\beta_3} \bar{g}_{\alpha_3}^{\beta_1}
+ \bar{g}_{\alpha_1}^{\beta_3} \bar{g}_{\alpha_2}^{\beta_1} \bar{g}_{\alpha_3}^{\beta_2}
+ \bar{g}_{\alpha_1}^{\beta_3}\bar{g}_{\alpha_2}^{\beta_2}\bar{g}_{\alpha_3}^{\beta_1} \Big)
\cr
 & + \frac{1}{15} \biggl[
  \bar{g}_{\alpha_1 \alpha_2} \Big(
  \bar{g}^{\beta_1 \beta_2} \bar{g}_{\alpha_3}^{\beta_3}
+ \bar{g}^{\beta_1 \beta_3} \bar{g}_{\alpha_3}^{\beta_2}
+ \bar{g}^{\beta_2 \beta_3}\bar{g}_{\alpha_3}^{\beta_1} \Big)
+ \bar{g}_{\alpha_1 \alpha_3} \Big(
  \bar{g}^{\beta_1 \beta_2}  \bar{g}_{\alpha_2}^{\beta_3}
+ \bar{g}^{\beta_1 \beta_3} \bar{g}_{\alpha_2}^{\beta_2}
+\bar{g}^{\beta_2 \beta_3}\bar{g}_{\alpha_2}^{\beta_1} \Big)
\cr
&\quad +
  \bar{g}_{\alpha_2 \alpha_3} \Big(
  \bar{g}^{\beta_1 \beta_2} \bar{g}_{\alpha_1}^{\beta_3}
+ \bar{g}^{\beta_1 \beta_3}\bar{g}_{\alpha_1}^{\beta_2}
+ \bar{g}^{\beta_2 \beta_3}\bar{g}_{\alpha_1}^{\beta_1} \Big)
\biggr],
\end{align}
\begin{align}
\Delta_{\alpha_1^{} \alpha_2^{} \alpha_3^{} \alpha_4^{}}^{\beta_1^{} \beta_2^{} \beta_3^{}\beta_4^{}}
(4,p) =
\frac{1}{576} \sum_{P(\alpha), P(\beta)}
\left(
\bar{g}_{\alpha_1^{}}^{\beta_1^{}} \bar{g}_{\alpha_2^{}}^{\beta_2^{}} 
\bar{g}_{\alpha_3^{}}^{\beta_3^{}} \bar{g}_{\alpha_4^{}}^{\beta_4^{}}
- \frac67 \bar{g}_{\alpha_1^{} \alpha_2^{}}^{} \bar{g}^{\beta_1^{}\beta_2^{}}
\bar{g}_{\alpha_3^{}}^{\beta_3^{}} \bar{g}_{\alpha_4^{}}^{\beta_4^{}}
+ \frac{3}{35} 
\bar{g}_{\alpha_1^{} \alpha_2^{}}^{} \bar{g}_{\alpha_3^{} \alpha_4^{}}^{}
\bar{g}^{\beta_1^{}\beta_2^{}} \bar{g}^{\beta_3^{}\beta_4^{}}
\right),
\end{align}
\begin{align}
& \Delta_{\alpha_1^{} \alpha_2^{} \alpha_3^{} \alpha_4^{} \alpha_5^{}}^{\beta_1^{} 
\beta_2^{} \beta_3^{}\beta_4^{} \beta_5^{}} (5,p)
\cr &
= - \left( \frac{1}{120} \right)^2 \sum_{P(\alpha),P(\beta)}
\left(
\bar{g}_{\alpha_1^{}}^{\beta_1^{}} \bar{g}_{\alpha_2^{}}^{\beta_2^{}}
\bar{g}_{\alpha_3^{}}^{\beta_3^{}} \bar{g}_{\alpha_4^{}}^{\beta_4^{}}
\bar{g}_{\alpha_5^{}}^{\beta_5^{}}
- \frac{10}{9}
\bar{g}_{\alpha_1^{}\alpha_2^{}}^{} \bar{g}^{\beta_1^{}\beta_2^{}}
\bar{g}_{\alpha_3^{}}^{\beta_3^{}} \bar{g}_{\alpha_4^{}}^{\beta_4^{}}
\theta_{\alpha_5^{}}^{\beta_5^{}}
+ \frac{5}{21}
\bar{g}_{\alpha_1^{}\alpha_2^{}}^{} \bar{g}_{\alpha_3^{}\alpha_4^{}}^{}
\bar{g}^{\beta_1^{}\beta_2^{}} \bar{g}^{\beta_3^{}\beta_4^{}}
\bar{g}_{\alpha_5^{}}^{\beta_5^{}}
\right),
\end{align}
\end{widetext}
where $\bar{g}_{\mu\nu}$ as
\begin{align}
\bar{g}_{\mu\nu}^{} =& g_{\mu\nu}^{} - \frac{1}{p^2} p_\mu^{} p_\nu^{}.
\end{align}

\vfil

\bibliographystyle{apsrev4-1}
\bibliography{ref}

\end{document}